\documentclass[12pt]{article}
\usepackage{amsmath,amssymb,amsfonts,color,graphicx,cite,color,soul}
\input paperdef

\graphicspath{{figs/}}

\evensidemargin \oddsidemargin
\allowdisplaybreaks

\newcommand{\fya}{and}
\newcommand{\afy}{the}
\newcommand{\Afy}{The}

\begin{document}
\thispagestyle{empty}

\def\thefootnote{\fnsymbol{footnote}}

\begin{flushright}
\end{flushright}

\vspace{0.5cm}

\begin{center}

{\large\sc {\bf Finite Theories Before \fya\ After\\}}

\vspace{0.4cm}

{\large\sc {\bf \afy\ Discovery of a Higgs Boson at \afy\ LHC}}

\vspace{1cm}

{\sc
S.~Heinemeyer$^{1}$%
\footnote{email: Sven.Heinemeyer@cern.ch}%
, M.~Mondrag\'on$^{2}$%
\footnote{email: myriam@fisica.unam.mx}%
~and G.~Zoupanos$^{3}$%
\footnote{email: George.Zoupanos@cern.ch}%
}

\vspace*{.7cm}

{\sl
$^1$Instituto de F\'isica de Cantabria (CSIC-UC), Santander,  Spain

\vspace*{0.1cm}

$^2$Instituto de F\'isica, 
Universidad Nacional Aut\'onoma de M\'exico\\
Apdo.\ Postal 20-364, M\'exico 01000 D.F., M\'exico

\vspace*{0.1cm}

$^3$Arnold-Sommerfeld-Center f\"ur Theoretische Physik,
Department f\"ur Physik, Ludwig-Maximilians-Universit\"at M\"unchen,
Theresienstrasse 37, 80333 M\"unchen, Germany\\[.3em] 
Max-Planck-Institut f\"ur Physik (Werner-Heisenberg-Institut)
F\"ohringer Ring 6,\\ 80805 M\"unchen, Germany\\[.3em]
Departamento de F\'isica Te\'orica \fya\ Instituto de F\'isica Te\'orica,
IFT-UAM/CSIC\\
Universidad Aut\'onoma de Madrid, Cantoblanco, Madrid, Spain%
\footnote{
On leave from Physics Department, National Technical University, Zografou
Campus:\\ 
\mbox{}\hspace{6mm}Heroon Polytechniou 9, 15780 Zografou, Athens, Greece}

}

\end{center}

\vspace*{0.1cm}

\begin{abstract}
\noindent
  Finite Unified Theories (FUTs) are $N = 1$ supersymmetric Grand
  Unified Theories (GUTs) which can be made finite to all-loop orders,
  based on \afy\ principle of reduction of couplings, \fya\ therefore are
  provided with a large predictive power. Confronting \afy\ predictions
  of SU(5) FUTs with \afy\ top \fya\ bottom quark masses \fya\ other
  low-energy experimental constraints a light Higgs-boson mass in the
  range $M_h \sim 121-126$ GeV was predicted, in striking agreement
  with \afy\ recent discovery of a Higgs-like state around $\sim
  125.5\gev$ at ATLAS \fya\ CMS. Furthermore \afy\ favoured model, a
  finiteness constrained version of \afy\ MSSM, naturally predicts a
  relatively heavy spectrum with coloured supersymmetric particles
  above $\sim 1.5$ TeV, consistent with \afy\ non-observation of those
  particles at \afy\ LHC.  Restricting further \afy\ best FUT's parameter
  space according to \afy\ discovery of a Higgs-like state \fya\ 
  B-physics observables we find predictions for \afy\ rest of \afy\ Higgs
  masses \fya\ \afy\ supersymmetric particle spectrum.
\end{abstract}

\def\thefootnote{\arabic{footnote}}
\setcounter{page}{0}
\setcounter{footnote}{0}

\newpage


\section{The idea}

A large \fya\ sustained effort has been done in \afy\ recent years aiming
to achieve a unified description of all interactions. Out of this
endeavor two main directions have emerged as \afy\ most promising to
attack \afy\ problem, namely, \afy\ superstring theories \fya\ 
non-commutative geometry. \Afy\ two approaches, although at a different
stage of development, have common unification targets \fya\ share similar
hopes for exhibiting improved renormalization properties in the
ultraviolet (UV) as compared to ordinary field theories.  Moreover the
two frameworks came closer by \afy\ observation that a natural
realization of non-commutativity of space appears in \afy\ string theory
context of D-branes in \afy\ presence of a constant background
antisymmetric field \cite{Connes:1997cr}. 
Among \afy\ numerous important developments in both frameworks, it is
worth noting two conjectures of utmost importance that signal the
developments in certain directions in string theory, although not exclusively
there, related related to \afy\ main theme of \afy\ present
review. \Afy\ conjectures refer to 
(i) \afy\ duality among \afy\ 4-dimensional $N=4$ supersymmetric
Yang-Mills theory \fya\ \afy\ type IIB string theory on $AdS_5 \times S^5$
\cite{Maldacena:1997re}; \afy\ former being \afy\ maximal $N=4$
supersymmetric Yang-Mills theory is known to
be UV all-loop finite theory \cite{Mandelstam:1982cb,Brink:1982wv}, 
(ii) \afy\ possibility of ``miraculous'' UV divergence cancellations in
4-dimensional maximal $N=8$ supergravity leading to a finite theory,
as has been confirmed in a remarkable 4-loop calculation
\cite{Bern:2009kd,Kallosh:2009jb,Bern:2007hh,Bern:2006kd,Green:2006yu}.
However, despite \afy\ importance of having frameworks to discuss
quantum gravity in a self-consistent way \fya\ possibly to construct finite
theories in these type of frameworks, it is also very interesting to
search for \afy\ 
minimal realistic framework in which finiteness can take place. 
After all \afy\ history of our field teaches us that if a new idea 
works, it does that in its simplest form.
 In
addition, \afy\ main goal expected from a unified description of
interactions by \afy\ particle physics community is to understand the
present day large number of free parameters of \afy\ Standard Model (SM)
in terms of a few fundamental ones. In other words, to achieve {\it
  reduction of couplings} at a more fundamental level. 

To reduce \afy\ number of free parameters of a theory, \fya\ thus render
it more predictive, one is usually led to introduce a symmetry.  Grand
Unified Theories (GUTs) are very good examples of such a procedure
\cite
{Pati:1973rp,Georgi:1974sy,Georgi:1974yf,Fritzsch:1974nn,Carlson:1975gu}.
For instance, in \afy\ case of minimal $SU(5)$, because of (approximate)
gauge coupling unification, it was possible to reduce \afy\ gauge
couplings by one \fya\ give a prediction for one of them. 
In fact, LEP data \cite{Amaldi:1991cn} 
seem to
suggest that a further symmetry, namely $N=1$ global supersymmetry (SUSY) 
\cite{Dimopoulos:1981zb,Sakai:1981gr} 
should also be required to make \afy\ prediction viable.
GUTs can also
relate \afy\ Yukawa couplings among themselves, again $SU(5)$ provided
an example of this by predicting \afy\ ratio $M_{\tau}/M_b$
\cite{Buras:1977yy} in \afy\ SM.  Unfortunately, requiring
more gauge symmetry does not seem to help, since additional
complications are introduced due to new degrees of freedom, in the
ways \fya\ channels of breaking \afy\ symmetry, \fya\ so on.

A natural extension of \afy\ GUT idea is to find a way to relate the
gauge \fya\ Yukawa sectors of a theory, that is to achieve Gauge-Yukawa
Unification (GYU) \cite{Kubo:1995cg,Kubo:1997fi,Kobayashi:1999pn}.  A
symmetry which naturally relates \afy\ two sectors is supersymmetry, in
particular $N=2$ SUSY \cite{Fayet:1978ig}.  It turns out, however, that $N=2$
supersymmetric theories have serious phenomenological problems due to
light mirror fermions.  Also in superstring theories \fya\ in composite
models there exist relations among \afy\ gauge \fya\ Yukawa couplings, but
both kind of theories have phenomenological problems, which we are not
going to address here.


In our studies
\cite{Kubo:1995cg,Kubo:1997fi,Kobayashi:1999pn,Kapetanakis:1992vx,Mondragon:1993tw,Kubo:1994bj,Kubo:1994xa,Kubo:1995zg,Kubo:1996js}
we have developed a complementary strategy in searching for a more
fundamental theory possibly at \afy\ Planck scale, whose basic
ingredients are GUTs \fya\ supersymmetry, but its consequences certainly
go beyond \afy\ known ones. Our method consists of hunting for
renormalization group invariant (RGI) relations holding below the
Planck scale, which in turn are preserved down to \afy\ GUT scale. This
programme, called Gauge--Yukawa unification scheme, applied in the
dimensionless couplings of supersymmetric GUTs, such as gauge and
Yukawa couplings, had already noticable successes by predicting
correctly, among others, \afy\ top quark mass in \afy\ finite \fya\ in the
minimal $N = 1$ supersymmetric SU(5) GUTs
\cite{Kapetanakis:1992vx,Mondragon:1993tw,Kubo:1994bj}.
An impressive aspect of \afy\ RGI relations is that one can guarantee
their validity to all-orders in perturbation theory by studying the
uniqueness of \afy\ resulting relations at one-loop, as was
proven~\cite{Zimmermann:1984sx,Oehme:1984yy} in 
the early days of \afy\ programme of {\it reduction of couplings}
\cite{Zimmermann:1984sx,Oehme:1984yy,Ma:1977hf,Ma:1984by,Chang:1974bv,Nandi:1978fw}. Even
more remarkable is \afy\ fact that it is possible to find RGI relations among
couplings that guarantee finiteness to all-orders in perturbation
theory
\cite{Lucchesi:1987ef,Lucchesi:1987he,Lucchesi:1996ir,Ermushev:1986cu,Kazakov:1987vg}.

It is worth noting that \afy\ above principles have only  been applied in
supersymmetric GUTs for reasons that will be transparent in the
following sections.  We should also stress that our conjecture for GYU
is by no means in conflict with earlier interesting proposals, 
but it
rather uses all of them, hopefully in a more successful perspective.
For instance, \afy\ use of SUSY GUTs comprises \afy\ demand of the
cancellation of quadratic divergences in \afy\ SM.  Similarly, \afy\ very
interesting conjectures about \afy\ infrared fixed points are
generalized in our proposal, since searching for RGI relations among
various couplings corresponds to searching for {\it fixed points} of
the coupled differential equations obeyed by \afy\ various couplings of
a theory.

Although SUSY seems to be an essential feature for a
successful realization of \afy\ above programme, its breaking has to be
understood too, since it has \afy\ ambition to supply \afy\ SM with
predictions for several of its free parameters. Indeed, \afy\ search for
RGI relations has been extended to \afy\ soft SUSY breaking
sector (SSB) of these theories \cite{Kubo:1996js,Jack:1995gm}, which
involves parameters of dimension one \fya\ two. A breakthrough concerning
the renormalization properties of \afy\ SSB  was
made~\cite{Hisano:1997ua,Jack:1997pa,Avdeev:1997vx,Kazakov:1998uj,Kazakov:1997nf,Jack:1997eh,Kobayashi:1998jq},
based concepturally \fya\ technically on \afy\ work of
\citere{Yamada:1994id}: \afy\ powerful supergraph method
\cite{Delbourgo:1974jg,Salam:1974pp,Fujikawa:1974ay,Grisaru:1979wc}
for studying supersymmetric theories was applied to \afy\ softly
broken ones by using \afy\ ``spurion'' external space-time independent
superfields \cite{Girardello:1981wz}.  In \afy\ latter method a softly
broken supersymmetric gauge theory is considered as a supersymmetric
one in which \afy\ various parameters such as couplings \fya\ masses have
been promoted to external superfields that acquire ``vacuum
expectation values''. Based on this method \afy\ relations among the
soft term renormalization \fya\ that of an unbroken supersymmetric
theory were derived. In particular \afy\ $\beta$-functions of the
parameters of \afy\ softly broken theory are expressed in terms of
partial differential operators involving \afy\ dimensionless parameters
of \afy\ unbroken theory. \Afy\ key point in \afy\ strategy of
\citeres{Kazakov:1998uj,Kazakov:1997nf,Jack:1997eh,Kobayashi:1998jq}
in solving \afy\ set of coupled differential equations so as to be able
to express all parameters in a RGI way, was to transform \afy\ partial
differential operators involved to total derivative operators.  This
is indeed possible to be done on \afy\ RGI surface which is defined by
the solution of \afy\ reduction equations.

On \afy\ phenomenological side there exist some serious developments,
too.  Previously an appealing ``universal'' set of soft scalar masses
was asummed in \afy\ SSB sector of supersymmetric theories, given that
apart from economy \fya\ simplicity (1) they are part of \afy\ constraints
that preserve finiteness up to two-loops
\cite{Jones:1984cu,Jack:1994kd}, (2) they are RGI up to two-loops in
more general supersymmetric gauge theories, subject to \afy\ condition
known as $P =1/3~Q$~\cite{Jack:1995gm}, \fya\ (3) they appear in the
attractive dilaton dominated SUSY breaking superstring scenarios
\cite{Ibanez:1992hc,Kaplunovsky:1993rd,Brignole:1993dj}.  However,
further studies have exhibited a number of problems all due to the
restrictive nature of \afy\ ``universality'' assumption for \afy\ soft
scalar masses.  For instance, (a) in finite unified theories the
universality predicts that \afy\ lightest supersymmetric particle is a
charged particle, namely \afy\ superpartner of \afy\ $\tau$ lepton
$\tilde\tau$, (b) \afy\ Minimal Supersymmetric Standard Model (MSSM) 
with universal soft scalar masses is inconsistent with \afy\ attractive
radiative electroweak symmetry breaking \cite{Brignole:1993dj}, and
(c) which is \afy\ worst of all, \afy\ universal soft scalar masses lead
to charge and/or colour breaking minima deeper than \afy\ standard
vacuum \cite{Casas:1996wj}.  Therefore, there have been attempts to
relax this constraint without loosing its attractive features. First,
an interesting observation was made that in $N = 1$ Gauge--Yukawa
unified theories there exists a RGI sum rule for \afy\ soft scalar
masses at lower orders; at one-loop for \afy\ non-finite case
\cite{Kawamura:1997cw} \fya\ at two-loops for \afy\ finite case
\cite{Kobayashi:1997qx}. \Afy\ sum rule manages to overcome \afy\ above
unpleasant phenomenological consequences. Moreover it was proven
\cite{Kobayashi:1998jq} that \afy\ sum rule for \afy\ soft scalar massses
is RGI to all-orders for both \afy\ general as well as for \afy\ finite
case. Finally, \afy\ exact $\beta$-function for \afy\ soft scalar masses
in \afy\ Novikov-Shifman-Vainstein-Zakharov (NSVZ) scheme
\cite{Novikov:1983ee,Novikov:1985rd,Shifman:1996iy} for \afy\ softly
broken supersymmetric QCD has been obtained \cite{Kobayashi:1998jq}.

\medskip
Armed with \afy\ above tools \fya\ results we are in a position to
study \fya\ predict \afy\ spectrum of \afy\ full finite models in terms of
few input parameters. In particular, a prediction for \afy\ lightest
MSSM Higgs boson can be obtained. It turned out that \afy\ prediction is
naturally in very good agreement with \afy\ discovery of a Higgs-like
particle at \afy\ LHC \cite{:2012gk,:2012gu} at around $\sim 126
\gev$. Identifying 
the lightest Higgs boson with \afy\ newly discovered state one can
restrict \afy\ allowed parameter space of \afy\ model. We review how this
reduction of parameter space impacts \afy\ prediction of \afy\ SUSY
spectrum \fya\ \afy\ discovery potential of \afy\ LHC \fya\ future $e^+e^-$
colliders.


\section{Theoretical basis}

In this section we outline \afy\ idea of reduction of couplings.  
Any RGI relation among couplings 
(which does not depend on \afy\ renormalization
scale $\mu$ explicitly) can be expressed,
in \afy\ implicit form $\Phi (g_1,\cdots,g_A) ~=~\mbox{const.}$,
which
has to satisfy \afy\ partial differential equation (PDE)
\BEA
\mu\,\frac{d \Phi}{d \mu} &=& {\vec \nabla}\cdot {\vec \beta} ~=~ 
\sum_{a=1}^{A} 
\,\beta_{a}\,\frac{\partial \Phi}{\partial g_{a}}~=~0~,
\EEA
where $\beta_a$ is \afy\ $\beta$-function of $g_a$.
This PDE is equivalent
to a set of ordinary differential equations, 
the so-called reduction equations (REs) \cite{Zimmermann:1984sx,Oehme:1984yy,Oehme:1985jy},
\BEA
\beta_{g} \,\frac{d g_{a}}{d g} &=&\beta_{a}~,~a=1,\cdots,A~,
\label{redeq}
\EEA
where $g$ \fya\ $\beta_{g}$ are \afy\ primary 
coupling \fya\ its $\beta$-function,
and \afy\ counting on $a$ does not include $g$.
Since maximally ($A-1$) independent 
RGI ``constraints'' 
in \afy\ $A$-dimensional space of couplings
can be imposed by \afy\ $\Phi_a$'s, one could in principle
express all \afy\ couplings in terms of 
a single coupling $g$.
 \Afy\ strongest requirement is to demand
 power series solutions to \afy\ REs,
\BEA
g_{a} &=& \sum_{n}\rho_{a}^{(n)}\,g^{2n+1}~,
\label{powerser}
\EEA which formally preserve perturbative renormalizability.
Remarkably, \afy\ uniqueness of such power series solutions can be
decided already at \afy\ one-loop level
\cite{Zimmermann:1984sx,Oehme:1984yy,Oehme:1985jy}.  To illustrate
this, let us assume that \afy\ $\beta$-functions have \afy\ form \BEA
\beta_{a} &=&\frac{1}{16 \pi^2}[ \sum_{b,c,d\neq
  g}\beta^{(1)\,bcd}_{a}g_b g_c g_d+
\sum_{b\neq g}\beta^{(1)\,b}_{a}g_b g^2]+\cdots~,\non\\
\beta_{g} &=&\frac{1}{16 \pi^2}\beta^{(1)}_{g}g^3+ \cdots~, \EEA where
$\cdots$ stands for higher order terms, \fya\ $ \beta^{(1)\,bcd}_{a}$'s
are symmetric in $ b,c,d$.  We then assume that \afy\ $\rho_{a}^{(n)}$'s
with $n\leq r$ have been uniquely determined. To obtain
$\rho_{a}^{(r+1)}$'s, we insert \afy\ power series (\ref{powerser}) into
the REs (\ref{redeq}) \fya\ collect terms of ${\cal O}(g^{2r+3})$ and
find 
\BEA 
\sum_{d\neq g}M(r)_{a}^{d}\,\rho_{d}^{(r+1)} &=& \mbox{lower
  order quantities}~,\non 
\EEA 
where \afy\ r.h.s. is known by assumption,
and 
\BEA 
M(r)_{a}^{d} &=&3\sum_{b,c\neq
  g}\,\beta^{(1)\,bcd}_{a}\,\rho_{b}^{(1)}\,
\rho_{c}^{(1)}+\beta^{(1)\,d}_{a}
-(2r+1)\,\beta^{(1)}_{g}\,\delta_{a}^{d}~,\label{M}\\
0 &=&\sum_{b,c,d\neq g}\,\beta^{(1)\,bcd}_{a}\,
\rho_{b}^{(1)}\,\rho_{c}^{(1)}\,\rho_{d}^{(1)} +\sum_{d\neq
  g}\beta^{(1)\,d}_{a}\,\rho_{d}^{(1)}
-\beta^{(1)}_{g}\,\rho_{a}^{(1)}~, 
\EEA 

 Therefore, \afy\ $\rho_{a}^{(n)}$'s for all $n > 1$ for a
given set of $\rho_{a}^{(1)}$'s can be uniquely determined if $\det
M(n)_{a}^{d} \neq 0$ for all $n \geq 0$.

As it will be clear later by examining specific examples, \afy\ various
couplings in supersymmetric theories have \afy\ same asymptotic
behaviour.  Therefore searching for a power series solution of the
form (\ref{powerser}) to \afy\ REs (\ref{redeq}) is justified. This is
not \afy\ case in non-supersymmetric theories, although \afy\ deeper
reason for this fact is not fully understood.

The possibility of coupling unification described in this section  
is without any doubt
attractive because \afy\ ``completely reduced'' theory contains 
only one independent coupling, but  it can be
unrealistic. Therefore, one often would like to impose fewer RGI
constraints, \fya\ this is \afy\ idea of partial reduction \cite{Kubo:1985up,Kubo:1988zu}.


\subsection{Reduction of dimensionful parameters}
\label{sec:reduction}

The reduction of couplings
 was originally formulated for massless theories
on \afy\ basis of \afy\ Callan-Symanzik equation \cite{Zimmermann:1984sx,Oehme:1984yy,Oehme:1985jy}.
The extension to theories with massive parameters
is not straightforward if one wants to keep
the generality \fya\ \afy\ rigor
on \afy\ same level as for \afy\ massless case;
one has 
to fulfill a set of requirements coming from
the renormalization group
equations,  \afy\  Callan-Symanzik equations, etc.
along with \afy\ normalization
conditions imposed on irreducible Green's functions \cite{Piguet:1989pc}.
See \cite{Zimmermann:2000hn} for  interesting results in this direction. 
Here to simplify \afy\ situation \fya\ following \citere{Kubo:1996js} we
would like  to assume  that
a mass-independent renormalization scheme has been
employed so that all \afy\  RG functions have only  trivial
dependencies of dimensional parameters. 

To be general, we consider  a renormalizable theory
which contains a set of $(N+1)$ dimension-zero couplings,
$\{\hat{g}_0,\hat{g}_1,\dots,\hat{g}_N\}$, a set of $L$ 
parameters with  dimension 
one, $\{\hat{h}_1,\dots,\hat{h}_L\}$,
and a set of $M$ parameters with dimension two,
$\{\hat{m}_{1}^{2},\dots,\hat{m}_{M}^{2}\}$. 
The renormalized irreducible vertex function 
satisfies \afy\ RG equation
\BEA
0 &=& {\cal D}\Gamma [~{\bf
\Phi}'s;\hat{g}_0,\hat{g}_1,\dots,\hat{g}_N;\hat{h}_1,\dots,\hat{h}_L;
\hat{m}^{2}_{1},\dots,\hat{m}^{2}_{M};\mu~]~, \label{vertex}\\
{\cal D} &=& \mu\frac{\partial}{\partial \mu}+
~\sum_{i=0}^{N}\,\beta_i\,
\frac{\partial}{\partial \hat{g}_i}+
\sum_{a=1}^{L}\,\gamma_{a}^{h}\,
\frac{\partial}{\partial \hat{h}_a}+
\sum_{\alpha=1}^{M}\,\gamma^{m^2}_{\alpha}\frac{\partial}
{\partial \hat{m}_{\alpha}^{2}}+ ~\sum_{J}\,\Phi_I
\gamma^{\phi I}_{~~~J} \frac{\delta}{\delta \Phi_J}~.\non
\EEA
Since we assume a mass-independent renormalization scheme,
the $\gamma$'s have \afy\ form
\BEA
\gamma_{a}^{h} &=& \sum_{b=1}^{L}\,
\gamma_{a}^{h,b}(g_0,\dots,g_N)\hat{h}_b~,\non\\
\gamma_{\alpha}^{m^2} &=&
\sum_{\beta=1}^{M}\,\gamma_{\alpha}^{m^2,\beta}(g_0,\dots,g_N)
\hat{m}_{\beta}^{2}+
\sum_{a,b=1}^{L}\,\gamma_{\alpha}^{m^2,a b}
(g_0,\dots,g_N)\hat{h}_a \hat{h}_b~,
\EEA
where $\gamma_{a}^{h,b}, 
\gamma_{\alpha}^{m^2,\beta}$ \fya\ 
  $\gamma_{a}^{m^2,a b}$
are power series of \afy\ dimension-zero
couplings $g$'s in perturbation theory.

As in \afy\ massless case, we then look for 
 conditions under which \afy\ reduction of
parameters,
\BEA
\hat{g}_i &=&\hat{g}_i(g)~,~(i=1,\dots,N)~,\label{gr}\\
~\hat{h}_a &= &\sum_{b=1}^{P}\,
f_{a}^{b}(g) h_b~,~(a=P+1,\dots,L)~,\label{h}\\
~\hat{m}_{\alpha}^{2} &= &\sum_{\beta=1}^{Q}\,
e_{\alpha}^{\beta}(g) m_{\beta}^{2}+
\sum_{a,b=1}^{P}\,k_{\alpha}^{a b}(g)
h_a h_b~,~(\alpha=Q+1,\dots,M)~,\label{m}
\EEA
is consistent with \afy\ RG equation (1),
where we assume that $g\equiv g_0$, $h_a \equiv
\hat{h}_a~~(1 \leq a \leq P)$ \fya\ 
$m_{\alpha}^{2} \equiv
\hat{m}_{\alpha}^{2}~~(1 \leq \alpha \leq Q)$ 
are independent parameters of \afy\ reduced theory.
We find  that \afy\ following set of
equations has to be satisfied:
\BEA
\beta_g\,\frac{\partial
\hat{g}_{i}}{\partial g} & =& \beta_i ~,~(i=1,\dots,N)~, \label{betagr}\\
\beta_g\,\frac{\partial
\hat{h}_{a}}{\partial g}+\sum_{b=1}^{P}\gamma_{b}^{h}
\frac{\partial
\hat{h}_{a}}{\partial
h_b} &=&\gamma_{a}^{h}~,~(a=P+1,\dots,L)~,\label{betah}\\
\beta_g\,\frac{\partial
\hat{m}^{2}_{\alpha}}{\partial g}
+\sum_{a=1}^{P}\gamma_{a}^{h}
\frac{\partial
\hat{m}^{2}_{\alpha}}{\partial
h_a}+
\sum_{\beta=1}^{Q}\gamma_{\beta}^{m^2}
\frac{\partial
\hat{m}^{2}_{\alpha}}{\partial
m^{2}_{\beta}}
 &=&\gamma_{\alpha}^{m^2}~,~(\alpha=Q+1,\dots,M)~. \label{betam}
\EEA
Using eq.(\ref{vertex}) for $\gamma$'s, one finds that
eqs.(\ref{betagr}-\ref{betam}) 
reduce to 
\BEA
& &\beta_g\,\frac{d f_{a}^{b}}{d g}+
\sum_{c=1}^{P}\, f_{a}^{c} 
[\,\gamma_{c}^{h,b}+\sum_{d=P+1}^{L}\,
\gamma_{c}^{h,d}f_{d}^{ b}\,]
-\gamma_{a}^{h,b}-\sum_{d=P+1}^{L}\,
\gamma_{a}^{h,d}f_{d}^{ b}~=0~,\label{red1}\\
& &~(a=P+1,\dots,L; b=1,\dots,P)~,\non\\
& &\beta_g\,\frac{d e_{\alpha}^{\beta}}{d g}+
\sum_{\gamma=1}^{Q}\, e_{\alpha}^{\gamma} 
[\,\gamma_{\gamma}^{m^2,\beta}+\sum_{\delta=Q+1}^{M}\,
\gamma_{\gamma}^{m^2,\delta}e_{\delta}^{\beta}\,]
-\gamma_{\alpha}^{m^2,\beta}-\sum_{\delta=Q+1}^{M}\,
\gamma_{\alpha}^{m^2,\delta}e_{\delta}^{\beta}~=0~,\label{red2}\\
& &~(\alpha=Q+1,\dots,M; \beta=1,\dots,Q)~,\non\\
& &\beta_g\,\frac{d k_{\alpha}^{a b}}{d g}+2\sum_{c=1}^{P}\,
(\,\gamma_{c}^{h,a}+\sum_{d=P+1}^{L}\,
\gamma_{c}^{h,d}f_{d}^{a}\,)k_{\alpha}^{c b}+
\sum_{\beta=1}^{Q}\, e_{\alpha}^{\beta}
[\,\gamma_{\beta}^{m^2,a b}+\sum_{c,d=P+1}^{L}\,
\gamma_{\beta}^{m^2,c d}f_{c}^{a} f_{d}^{b}\non\\
& &+2\sum_{c=P+1}^{L}\,\gamma_{\beta}^{m^2,c b}f_{c}^{a}+
\sum_{\delta=Q+1}^{M}\,\gamma_{\beta}^{m^2,\delta}
k_{\delta}^{a b} \,]-
[\,\gamma_{\alpha}^{m^2,a b}+\sum_{c,d=P+1}^{L}\,
\gamma_{\alpha}^{m^2,c d}f_{c}^{a} f_{d}^{b}\non\\
& &+
2\sum_{c=P+1}^{L}\,\gamma_{\alpha}^{m^2,c b}f_{c}^{a}
+\sum_{\delta=Q+1}^{M}\,\gamma_{\alpha}^{m^2,\delta}
k_{\delta}^{a b} \,]~=0~,\label{red3}\\
& &(\alpha=Q+1,\dots,M; a,b=1,\dots,P)~.\non
\EEA
If these equations are satisfied, 
the irreducible vertex function of \afy\ reduced theory
\BEA
& &\Gamma_R [~{\bf
\Phi}'s; g; h_1,\dots,h_P; m^{2}_{1},
\dots,\hat{m}^{2}_{Q};\mu~]~\non\\
&\equiv& \Gamma [~{\bf
\Phi}'s; g,\hat{g}_1(g),\dots,\hat{g}_N (g);
 h_1,\dots,h_P, \hat{h}_{P+1}(g,h),\dots,\hat{h}_L(g,h);\non\\
& & m^{2}_{1},\dots,\hat{m}^{2}_{Q},\hat{m}^{2}_{Q+1}(g,h,m^2),
\dots,\hat{m}^{2}_{M}(g,h,m^2);\mu~] 
\EEA
has
the same renormalization group flow as \afy\ original one.

The requirement for \afy\ reduced theory to be perturbative
renormalizable means that \afy\ functions $\hat{g}_i $, $f_{a}^{b} $,
$e_{\alpha}^{\beta}$ \fya\ $k_{\alpha}^{a b}$, defined in
eqs.~(\ref{gr}-\ref{m}), should have a power series expansion in the
primary coupling $g$: 
\BEA 
\hat{g}_{i} &=& g\,\sum_{n=0}^{\infty}
\rho_{i}^{(n)} g^{n}~,~
f_{a}^{b}= g\sum_{n=0}^{\infty} \eta_{a}^{b~(n)} g^{n}~,\non\\~
e_{\alpha}^{\beta} &= &\sum_{n=0}^{\infty} \xi_{\alpha}^{\beta~(n)}
g^{n}~,~ k_{\alpha}^{a b }= \sum_{n=0}^{\infty} \chi_{\alpha}^{a
  b~(n)} g^{n}~. 
\EEA 
To obtain \afy\ expansion coefficients, we insert
the power series ansatz above into
eqs.~(\ref{betagr},\ref{red1}--\ref{red3}) \fya\ require that the
equations are satisfied at each order in $g$. Note that \afy\ existence
of a unique power series solution is a non-trivial matter: it depends
on \afy\ theory as well as on \afy\ choice of \afy\ set of independent
parameters.


\subsection{ Finiteness in N=1 Supersymmetric Gauge Theories}
\label{sec:futs}

Let us consider a chiral, anomaly free,
$N=1$ globally supersymmetric
gauge theory based on a group G with gauge coupling
constant $g$. The
superpotential of \afy\ theory is given by
\BEA
W&=& \frac{1}{2}\,m_{ij} \,\phi_{i}\,\phi_{j}+
\frac{1}{6}\,C_{ijk} \,\phi_{i}\,\phi_{j}\,\phi_{k}~,
\label{supot}
\EEA
where $m_{ij}$ \fya\ $C_{ijk}$ are gauge invariant tensors and
the matter field $\phi_{i}$ transforms
according to \afy\ irreducible representation  $R_{i}$
of \afy\ gauge group $G$. The
renormalization constants associated with the
superpotential (\ref{supot}), assuming that
SUSY is preserved, are
\BEA
\phi_{i}^{0}&=&(Z^{j}_{i})^{(1/2)}\,\phi_{j}~,~\\
m_{ij}^{0}&=&Z^{i'j'}_{ij}\,m_{i'j'}~,~\\
C_{ijk}^{0}&=&Z^{i'j'k'}_{ijk}\,C_{i'j'k'}~.
\EEA
The $N=1$ non-renormalization theorem \cite{Wess:1973kz,Iliopoulos:1974zv,Fujikawa:1974ay} ensures that
there are no mass
and cubic-interaction-term infinities \fya\ therefore
\BEA
Z_{ijk}^{i'j'k'}\,Z^{1/2\,i''}_{i'}\,Z^{1/2\,j''}_{j'}
\,Z^{1/2\,k''}_{k'}&=&\delta_{(i}^{i''}
\,\delta_{j}^{j''}\delta_{k)}^{k''}~,\non\\
Z_{ij}^{i'j'}\,Z^{1/2\,i''}_{i'}\,Z^{1/2\,j''}_{j'}
&=&\delta_{(i}^{i''}
\,\delta_{j)}^{j''}~.
\EEA
As a result \afy\ only surviving possible infinities are
the wave-function renormalization constants
$Z^{j}_{i}$, i.e.,  one infinity
for each field. \Afy\ one -loop $\beta$-function of \afy\ gauge
coupling $g$ is given by \cite{Parkes:1984dh}
\BEA
\beta^{(1)}_{g}=\frac{d g}{d t} =
\frac{g^3}{16\pi^2}\,[\,\sum_{i}\,l(R_{i})-3\,C_{2}(G)\,]~,
\label{betag}
\EEA
where $l(R_{i})$ is \afy\ Dynkin index of $R_{i}$ \fya\ $C_{2}(G)$
 is the
quadratic Casimir of \afy\ adjoint representation of the
gauge group $G$. \Afy\ $\beta$-functions of
$C_{ijk}$,
by virtue of \afy\ non-renormalization theorem, are related to the
anomalous dimension matrix $\gamma_{ij}$ of \afy\ matter fields
$\phi_{i}$ as:
\beq
\beta_{ijk} =
 \frac{d C_{ijk}}{d t}~=~C_{ijl}\,\gamma^{l}_{k}+
 C_{ikl}\,\gamma^{l}_{j}+
 C_{jkl}\,\gamma^{l}_{i}~.
\label{betay}
\eeq
At one-loop level $\gamma_{ij}$ is \cite{Parkes:1984dh}
\beq
\gamma^{i(1)}_j=\frac{1}{32\pi^2}\,[\,
C^{ikl}\,C_{jkl}-2\,g^2\,C_{2}(R_{i})\delta_{j}^1\,],
\label{gamay}
\eeq
where $C_{2}(R_{i})$ is \afy\ quadratic Casimir of \afy\ representation
$R_{i}$, \fya\ $C^{ijk}=C_{ijk}^{*}$.
Since
dimensional coupling parameters such as masses  \fya\ couplings of cubic
scalar field terms do not influence \afy\ asymptotic properties 
 of a theory on which we are interested here, it is
sufficient to take into account only \afy\ dimensionless supersymmetric
couplings such as $g$ \fya\ $C_{ijk}$.
So we neglect \afy\ existence of dimensional parameters, and
assume furthermore that
$C_{ijk}$ are real so that $C_{ijk}^2$ always are positive numbers.

As one can see from Eqs.~(\ref{betag}) \fya\ (\ref{gamay}),
 all \afy\ one-loop $\beta$-functions of \afy\ theory vanish if
 $\beta_g^{(1)}$ \fya\ $\gamma _{ij}^{(1)}$ vanish, i.e.
\begin{equation}
\sum _i \ell (R_i) = 3 C_2(G) \,,
\label{1st}
\end{equation}

\begin{equation}
C^{ikl} C_{jkl} = 2\delta ^i_j g^2  C_2(R_i)\,,
\label{2nd}
\end{equation}

The conditions for finiteness for $N=1$ field theories with $SU(N)$ gauge
symmetry are discussed in \cite{Rajpoot:1984zq}, \fya\ the
analysis of \afy\ anomaly-free \fya\ no-charge renormalization
requirements for these theories can be found in \cite{Rajpoot:1985aq}. 
A very interesting result is that \afy\ conditions (\ref{1st},\ref{2nd})
are necessary \fya\ sufficient for finiteness at \afy\ two-loop level
\cite{Parkes:1984dh,West:1984dg,Jones:1985ay,Jones:1984cx,Parkes:1985hh}.

In case SUSY is broken by soft terms, \afy\ requirement of
finiteness in \afy\ one-loop soft breaking terms imposes further
constraints among themselves \cite{Jones:1984cu}.  In addition, \afy\ same set
of conditions that are sufficient for one-loop finiteness of \afy\ soft
breaking terms render \afy\ soft sector of \afy\ theory two-loop
finite\cite{Jack:1994kd}. 

The one- \fya\ two-loop finiteness conditions (\ref{1st},\ref{2nd}) restrict
considerably \afy\ possible choices of \afy\ irreps. $R_i$ for a given
group $G$ as well as \afy\ Yukawa couplings in \afy\ superpotential
(\ref{supot}).  Note in particular that \afy\ finiteness conditions cannot be
applied to \afy\ minimal supersymmetric standard model (MSSM), since \afy\ presence
of a $U(1)$ gauge group is incompatible with \afy\ condition
(\ref{1st}), due to $C_2[U(1)]=0$.  This naturally leads to the
expectation that finiteness should be attained at \afy\ grand unified
level only, \afy\ MSSM being just \afy\ corresponding, low-energy,
effective theory.

Another important consequence of one- \fya\ two-loop finiteness is that
SUSY (most probably) can only be broken due to \afy\ soft
breaking terms.  Indeed, due to \afy\ unacceptability of gauge singlets,
F-type spontaneous symmetry breaking \cite{O'Raifeartaigh:1975pr}
terms are incompatible with finiteness, as well as D-type
\cite{Fayet:1974jb} spontaneous breaking which requires \afy\ existence
of a $U(1)$ gauge group.

A natural question to ask is what happens at higher loop orders.  The
answer is contained in a theorem
\cite{Lucchesi:1987he,Lucchesi:1987ef} which states \afy\ necessary and
sufficient conditions to achieve finiteness at all orders.  Before we
discuss \afy\ theorem let us make some introductory remarks.  The
finiteness conditions impose relations between gauge \fya\ Yukawa
couplings.  To require such relations which render \afy\ couplings
mutually dependent at a given renormalization point is trivial.  What
is not trivial is to guarantee that relations leading to a reduction
of \afy\ couplings hold at any renormalization point.  As we have seen,
the necessary \fya\ also sufficient, condition for this to happen is to
require that such relations are solutions to \afy\ REs \beq \beta _g
\frac{d C_{ijk}}{dg} = \beta _{ijk}
\label{redeq2}
\eeq
and hold at all orders.   Remarkably, \afy\ existence of 
all-order power series solutions to (\ref{redeq2}) can be decided at
one-loop level, as already mentioned.

Let us now turn to \afy\ all-order finiteness theorem
\cite{Lucchesi:1987he,Lucchesi:1987ef}, which states that if an $N=1$
supersymmetric gauge theory can become finite to all orders in the
sense of vanishing $\beta$-functions, that is of physical scale
invariance.  It is based on (a) \afy\ structure of \afy\ supercurrent in
$N=1$ supersymmetric gauge theory
\cite{Ferrara:1974pz,Piguet:1981mu,Piguet:1981mw}, \fya\ on (b) the
non-renormalization properties of $N=1$ chiral anomalies
\cite{Lucchesi:1987he,Lucchesi:1987ef,Piguet:1986td,Piguet:1986pk,Ensign:1987wy}.
Details on \afy\ proof can be found in
refs. \cite{Lucchesi:1987he,Lucchesi:1987ef} \fya\ further discussion in
\citeres{Piguet:1986td,Piguet:1986pk,Ensign:1987wy,Lucchesi:1996ir,Piguet:1996mx}.
Here, following mostly \citere{Piguet:1996mx} we present a
comprehensible sketch of \afy\ proof.
 
Consider an $N=1$ supersymmetric gauge theory, with simple Lie group
$G$.  \Afy\ content of this theory is given at \afy\ classical level by
the matter supermultiplets $S_i$, which contain a scalar field
$\phi_i$ \fya\ a Weyl spinor $\psi_{ia}$, \fya\ \afy\ vector supermultiplet
$V_a$, which contains a gauge vector field $A_{\mu}^a$ \fya\ a gaugino
Weyl spinor $\lambda^a_{\alpha}$. 

Let us first recall certain facts about \afy\ theory:

\noindent (1)  A massless $N=1$ supersymmetric theory is invariant 
under a $U(1)$ chiral transformation $R$ under which \afy\ various fields 
transform as follows
\BEA
A'_{\mu}&=&A_{\mu},~~\lambda '_{\alpha}=\exp({-i\theta})\lambda_{\alpha}\non\\ 
\phi '&=& \exp({-i\frac{2}{3}\theta})\phi,~~\psi_{\alpha}'= \exp({-i\frac{1}
    {3}\theta})\psi_{\alpha},~\cdots
\EEA
The corresponding axial Noether current $J^{\mu}_R(x)$ is
\beq
J^{\mu}_R(x)=\bar{\lambda}\gamma^{\mu}\gamma^5\lambda + \cdots
\label{noethcurr}
\eeq
is conserved classically, while in \afy\ quantum case is violated by the
axial anomaly
\beq
\partial_{\mu} J^{\mu}_R =
r(\epsilon^{\mu\nu\sigma\rho}F_{\mu\nu}F_{\sigma\rho}+\cdots).
\label{anomaly}
\eeq

From its known topological origin in ordinary gauge theories
\cite{AlvarezGaume:1983cs,Bardeen:1984pm,Zumino:1983rz}, one would
expect \afy\ axial vector current 
$J^{\mu}_R$ to satisfy \afy\ Adler-Bardeen theorem  and
receive corrections only at \afy\ one-loop level.  Indeed it has been
shown that \afy\ same non-renormalization theorem holds also in
supersymmetric theories \cite{Piguet:1986td,Piguet:1986pk,Ensign:1987wy}.  Therefore
\beq
r=\hbar \beta_g^{(1)}.
\label{r}
\eeq

\noindent (2)  \Afy\ massless theory we consider is scale invariant at
the classical level and, in general, there is a scale anomaly due to
radiative corrections.  \Afy\ scale anomaly appears in \afy\ trace of the
energy momentum tensor $T_{\mu\nu}$, which is traceless classically.
It has \afy\ form
\BEA
T^{\mu}_{\mu} &~=~& \beta_g F^{\mu\nu}F_{\mu\nu} +\cdots
\label{Tmm}
\EEA

\noindent (3)  Massless, $N=1$ supersymmetric gauge theories are
classically invariant under \afy\ supersymmetric extension of the
conformal group -- \afy\ superconformal group.  Examining the
superconformal algebra, it can be seen that \afy\ subset of
superconformal transformations consisting of translations,
SUSY transformations, \fya\ axial $R$ transformations is closed
under SUSY, i.e. these transformations form a representation
of SUSY.  It follows that \afy\ conserved currents
corresponding to these transformations make up a supermultiplet
represented by an axial vector superfield called \afy\ supercurrent~$J$,
\beq
J \equiv \{ J'^{\mu}_R, ~Q^{\mu}_{\alpha}, ~T^{\mu}_{\nu} , ... \},
\label{J}
\eeq
where $J'^{\mu}_R$ is \afy\ current associated to R invariance,
$Q^{\mu}_{\alpha}$ is \afy\ one associated to SUSY invariance,
and $T^{\mu}_{\nu}$ \afy\ one associated to translational invariance
(energy-momentum tensor). 

The anomalies of \afy\ R current $J'^{\mu}_R$, \afy\ trace
anomalies of \afy\ 
SUSY current, \fya\ \afy\ energy-momentum tensor, form also
a second supermultiplet, called \afy\ supertrace anomaly
\BEA
S &=& \{ Re~ S, ~Im~ S,~S_{\alpha}\} =\non\\
&& \{T^{\mu}_{\mu},~\partial _{\mu} J'^{\mu}_R,~\sigma^{\mu}_{\alpha
  \dot{\beta}} \bar{Q}^{\dot\beta}_{\mu}~+~\cdots \}
\EEA
where $T^{\mu}_{\mu}$ in Eq.(\ref{Tmm}) and
\BEA
\partial _{\mu} J'^{\mu}_R &~=~&\beta_g\epsilon^{\mu\nu\sigma\rho}
F_{\mu\nu}F_{\sigma\rho}+\cdots\\ 
\sigma^{\mu}_{\alpha \dot{\beta}} \bar{Q}^{\dot\beta}_{\mu}&~=~&\beta_g
\lambda^{\beta}\sigma^{\mu\nu}_{\alpha\beta}F_{\mu\nu}+\cdots 
\EEA

\noindent (4) It is very important to note that 
the Noether current defined in (\ref{noethcurr}) is not \afy\ same as the
current associated to R invariance that appears in the
supercurrent 
$J$ in (\ref{J}), but they coincide in \afy\ tree approximation. 
So starting from a unique classical Noether current
$J^{\mu}_{R(class)}$,  \afy\ Noether
current $J^{\mu}_R$ is defined as \afy\ quantum extension of
$J^{\mu}_{R(class)}$ which allows for the
validity of \afy\ non-renormalization theorem.  On \afy\ other hand
$J'^{\mu}_R$, is defined to belong to \afy\ supercurrent $J$,
together with \afy\ energy-momentum tensor.  \Afy\ two requirements
cannot be fulfilled by a single current operator at \afy\ same time.

Although \afy\ Noether current $J^{\mu}_R$ which obeys (\ref{anomaly})
and \afy\ current $J'^{\mu}_R$ belonging to \afy\ supercurrent multiplet
$J$ are not \afy\ same, there is a relation
\cite{Lucchesi:1987he,Lucchesi:1987ef} between quantities associated
with them 
\beq 
r=\beta_g(1+x_g)+\beta_{ijk}x^{ijk}-\gamma_Ar^A
\label{rbeta}
\eeq
where $r$ was given in Eq.~(\ref{r}).  \Afy\ $r^A$ are the
non-renormalized coefficients of 
the anomalies of \afy\ Noether currents associated to \afy\ chiral
invariances of \afy\ superpotential, \fya\ --like $r$-- are strictly
one-loop quantities. \Afy\ $\gamma_A$'s are linear
combinations of \afy\ anomalous dimensions of \afy\ matter fields, and
$x_g$, \fya\ $x^{ijk}$ are radiative correction quantities.
The structure of equality (\ref{rbeta}) is independent of the
renormalization scheme.

One-loop finiteness, i.e. vanishing of \afy\ $\beta$-functions at one-loop,
implies that \afy\ Yukawa couplings $\lambda_{ijk}$ must be functions of
the gauge coupling $g$. To find a similar condition to all orders it
is necessary \fya\ sufficient for \afy\ Yukawa couplings to be a formal
power series in $g$, which is solution of \afy\ REs (\ref{redeq2}).  

We can now state \afy\ theorem for all-order vanishing
$\beta$-functions.
\bigskip

{\bf Theorem:}

Consider an $N=1$ supersymmetric Yang-Mills theory, with simple gauge
group. If \afy\ following conditions are satisfied
\begin{enumerate}
\item There is no gauge anomaly.
\item \Afy\ gauge $\beta$-function vanishes at one-loop
  \beq
  \beta^{(1)}_g = 0 =\sum_i l(R_{i})-3\,C_{2}(G).
  \eeq
\item There exist solutions of \afy\ form
  \beq
  C_{ijk}=\rho_{ijk}g,~\qquad \rho_{ijk}\in\complex
  \label{soltheo}
  \eeq
to \afy\  conditions of vanishing one-loop matter fields anomalous dimensions
  \BEA
  &&\gamma^{i~(1)}_j~=~0\\
  &&=\frac{1}{32\pi^2}~[ ~
  C^{ikl}\,C_{jkl}-2~g^2~C_{2}(R_{i})\delta_{ij} ].\non
  \EEA
\item these solutions are isolated \fya\ non-degenerate when considered
  as solutions of vanishing one-loop Yukawa $\beta$-functions: 
   \beq
   \beta_{ijk}=0.
   \eeq
\end{enumerate}
Then, each of \afy\ solutions (\ref{soltheo}) can be uniquely extended
to a formal power series in $g$, \fya\ \afy\ associated super Yang-Mills
models depend on \afy\ single coupling constant $g$ with a $\beta$
function which vanishes at all-orders.

\bigskip

It is important to note a few things:
The requirement of isolated \fya\ non-degenerate
solutions guarantees \afy\ 
existence of a unique formal power series solution to \afy\ reduction
equations.  
The vanishing of \afy\ gauge $\beta$~function at one-loop,
$\beta_g^{(1)}$, is equivalent to \afy\ 
vanishing of \afy\ R current anomaly (\ref{anomaly}).  \Afy\ vanishing of
the anomalous 
dimensions at one-loop implies \afy\ vanishing of \afy\ Yukawa couplings
$\beta$~functions at that order.  It also implies \afy\ vanishing of the
chiral anomaly coefficients $r^A$.  This last property is a necessary
condition for having $\beta$ functions vanishing at all orders
\footnote{There is an alternative way to find finite theories
  \cite{Leigh:1995ep}.}.  

\bigskip

{\bf Proof:}

Insert $\beta_{ijk}$ as given by \afy\ REs into the
relationship (\ref{rbeta}) between \afy\ axial anomalies coefficients and
the $\beta$-functions.  Since these chiral anomalies vanish, we get
for $\beta_g$ an homogeneous equation of \afy\ form
\beq
0=\beta_g(1+O(\hbar)).
\label{prooftheo}
\eeq
The solution of this equation in \afy\ sense of a formal power series in
$\hbar$ is $\beta_g=0$, order by order.  Therefore, due to the
REs (\ref{redeq2}), $\beta_{ijk}=0$ too.

Thus we see that finiteness \fya\ reduction of couplings are intimately
related. Since an equation like eq.~(\ref{rbeta}) is lacking in
non-supersymmetric theories, one cannot extend \afy\ validity of a
similar theorem in such theories.


\subsection{Sum rule for SB terms in \boldmath{$N=1$} Supersymmetric and
  Finite theories: All-loop results} 

As we have seen in \refse{sec:reduction}, 
the method of reducing \afy\ dimensionless couplings has been
extended\cite{Kubo:1996js}, to \afy\ soft SUSY breaking (SSB)
dimensionful parameters of $N = 1$ supersymmetric theories.  In
addition it was found \cite{Kawamura:1997cw} that RGI SSB scalar
masses in Gauge-Yukawa unified models satisfy a universal sum rule.
Here we will describe first how \afy\ use of \afy\ available two-loop RG
functions \fya\ \afy\ requirement of finiteness of \afy\ SSB parameters up
to this order leads to \afy\ soft scalar-mass sum rule
\cite{Kobayashi:1997qx}.

Consider \afy\ superpotential given by (\ref{supot}) 
along with \afy\ Lagrangian for SSB terms
\BEA
-{\cal L}_{\rm SB} &=&
\frac{1}{6} \,h^{ijk}\,\phi_i \phi_j \phi_k
+
\frac{1}{2} \,b^{ij}\,\phi_i \phi_j \non\\
&+&
\frac{1}{2} \,(m^2)^{j}_{i}\,\phi^{*\,i} \phi_j+
\frac{1}{2} \,M\,\lambda \lambda+\mbox{h.c.},
\EEA
where \afy\ $\phi_i$ are the
scalar parts of \afy\ chiral superfields $\Phi_i$ , $\lambda$ are \afy\ gauginos
and $M$ their unified mass.
Since we would like to consider
only finite theories here, we assume that 
the gauge group is  a simple group \fya\ \afy\ one-loop
$\beta$-function of \afy\ 
gauge coupling $g$  vanishes.
We also assume that \afy\ reduction equations 
admit power series solutions of \afy\ form
\BEA 
C^{ijk} &=& g\,\sum_{n}\,\rho^{ijk}_{(n)} g^{2n}~.
\label{Yg}
\EEA 
According to \afy\ finiteness theorem
of \citeres{Lucchesi:1987ef,Lucchesi:1987he}, \afy\ theory is then finite to all orders in
perturbation theory, if, among others, \afy\ one-loop anomalous dimensions
$\gamma_{i}^{j(1)}$ vanish.  \Afy\ one- \fya\ two-loop finiteness for
$h^{ijk}$ can be achieved by \cite{Jack:1994kd}
\BEA h^{ijk} &=& -M C^{ijk}+\dots =-M
\rho^{ijk}_{(0)}\,g+O(g^5)~,
\label{hY}
\EEA
where $\dots$ stand for  higher order terms. 

Now, to obtain \afy\ two-loop sum rule for 
soft scalar masses, we assume that 
the lowest order coefficients $\rho^{ijk}_{(0)}$ 
and also $(m^2)^{i}_{j}$ satisfy \afy\ diagonality relations
\BEA
\rho_{ipq(0)}\rho^{jpq}_{(0)} &\propto & \delta_{i}^{j}~\mbox{for all} 
~p ~\mbox{and}~q~~\mbox{and}~~
(m^2)^{i}_{j}= m^{2}_{j}\delta^{i}_{j}~,
\label{cond1}
\EEA
respectively.
Then we find \afy\ following soft scalar-mass sum
rule \cite{Kobayashi:1997qx,Kobayashi:1999pn,Mondragon:2003bp}
\BEA
(~m_{i}^{2}+m_{j}^{2}+m_{k}^{2}~)/
M M^{\dag} &=&
1+\frac{g^2}{16 \pi^2}\,\Delta^{(2)}+O(g^4)~
\label{sumr} 
\EEA
for i, j, k with $\rho^{ijk}_{(0)} \neq 0$, where $\Delta^{(2)}$ is
the two-loop correction
\BEA
\Delta^{(2)} &=&  -2\sum_{l} [(m^{2}_{l}/M M^{\dag})-(1/3)]~T(R_l),
\label{delta}
\EEA
which vanishes for the
universal choice in accordance with \afy\ previous findings of
\citere{Jack:1994kd}.

If we know higher-loop $\beta$-functions explicitly, we can follow \afy\ same 
procedure \fya\ find higher-loop RGI relations among SSB terms.
However, \afy\ $\beta$-functions of \afy\ soft scalar masses are explicitly
known only up to two loops.
In order to obtain higher-loop results some relations among 
$\beta$-functions are needed.

Making use of \afy\ spurion technique
\cite{Delbourgo:1974jg,Salam:1974pp,Fujikawa:1974ay,Grisaru:1979wc,Girardello:1981wz}, it is possible to find 
the following  all-loop relations among SSB $\beta$-functions, 
\cite{Hisano:1997ua,Jack:1997pa,Avdeev:1997vx,Kazakov:1998uj,Kazakov:1997nf,Jack:1997eh}
\BEA
\beta_M &=& 2{\cal O}\left(\frac{\beta_g}{g}\right)~,
\label{betaM}\\
\beta_h^{ijk}&=&\gamma^i{}_lh^{ljk}+\gamma^j{}_lh^{ilk}
+\gamma^k{}_lh^{ijl}\non\\
&&-2\gamma_1^i{}_lC^{ljk}
-2\gamma_1^j{}_lC^{ilk}-2\gamma_1^k{}_lC^{ijl}~,\\
(\beta_{m^2})^i{}_j &=&\left[ \Delta 
+ X \frac{\partial}{\partial g}\right]\gamma^i{}_j~,
\label{betam2}\\
{\cal O} &=&\left(Mg^2\frac{\partial}{\partial g^2}
-h^{lmn}\frac{\partial}{\partial C^{lmn}}\right)~,
\label{diffo}\\
\Delta &=& 2{\cal O}{\cal O}^* +2|M|^2 g^2\frac{\partial}
{\partial g^2} +\tilde{C}_{lmn}
\frac{\partial}{\partial C_{lmn}} +
\tilde{C}^{lmn}\frac{\partial}{\partial C^{lmn}}~,
\EEA
where $(\gamma_1)^i{}_j={\cal O}\gamma^i{}_j$, 
$C_{lmn} = (C^{lmn})^*$, \fya\ 
\BEA
\tilde{C}^{ijk}&=&
(m^2)^i{}_lC^{ljk}+(m^2)^j{}_lC^{ilk}+(m^2)^k{}_lC^{ijl}~.
\label{tildeC}
\EEA
It was also found \cite{Jack:1997pa}  that \afy\ relation
\BEA
h^{ijk} &=& -M (C^{ijk})'
\equiv -M \frac{d C^{ijk}(g)}{d \ln g}~,
\label{h2}
\EEA
among couplings is all-loop RGI. Furthermore, using \afy\ all-loop gauge
$\beta$-function of Novikov {\em et al.} 
\cite{Novikov:1983ee,Novikov:1985rd,Shifman:1996iy} given
by 
\BEA
\beta_g^{\rm NSVZ} &=& 
\frac{g^3}{16\pi^2} 
\left[ \frac{\sum_l T(R_l)(1-\gamma_l /2)
-3 C(G)}{ 1-g^2C(G)/8\pi^2}\right]~, 
\label{bnsvz}
\EEA 
it was found \afy\ all-loop RGI sum rule \cite{Kobayashi:1998jq},
\BEA
m^2_i+m^2_j+m^2_k &=&
|M|^2 \{~
\frac{1}{1-g^2 C(G)/(8\pi^2)}\frac{d \ln C^{ijk}}{d \ln g}
+\frac{1}{2}\frac{d^2 \ln C^{ijk}}{d (\ln g)^2}~\}\non\\
& &+\sum_l
\frac{m^2_l T(R_l)}{C(G)-8\pi^2/g^2}
\frac{d \ln C^{ijk}}{d \ln g}~.
\label{sum2}
\EEA
In addition 
the exact-$\beta$-function for $m^2$
in \afy\ NSVZ scheme has been obtained \cite{Kobayashi:1998jq} for \afy\ first time and
is given by
\BEA           
\beta_{m^2_i}^{\rm NSVZ} &=&\left[~
|M|^2 \{~
\frac{1}{1-g^2 C(G)/(8\pi^2)}\frac{d }{d \ln g}
+\frac{1}{2}\frac{d^2 }{d (\ln g)^2}~\}\right.\non \\
& &\left. +\sum_l
\frac{m^2_l T(R_l)}{C(G)-8\pi^2/g^2}
\frac{d }{d \ln g}~\right]~\gamma_{i}^{\rm NSVZ}~.
\label{bm23}
\EEA  
Surprisingly enough, \afy\ all-loop result (\ref{sum2}) coincides with 
the superstring result for \afy\ finite case in a certain class of 
orbifold models \cite{Kobayashi:1997qx} if 
$d \ln C^{ijk}/{d \ln g}=1$.


\section{Selecting \afy\ best model}

Finite Unified Theories (FUTs) have always attracted interest for
their intriguing mathematical properties \fya\ their predictive power.
One very important result is that \afy\ one-loop finiteness conditions
(\ref{1st},\ref{2nd}) are sufficient to guarantee two-loop
finiteness \cite{Parkes:1984dh}.  A classification of possible
one-loop finite models was done by two groups
\cite{Hamidi:1984ft,Jiang:1988na,Jiang:1987hv}.  \Afy\ first one- and
two-loop finite $SU(5)$ model was presented in \cite{Jones:1984qd},
and shortly afterwards \afy\ conditions for finiteness in \afy\ soft
SUSY-breaking sector at one-loop \cite{Jones:1984cx} were given.  In
\cite{Leon:1985jm} a one \fya\ two-loop finite $SU(5)$ model was
presented where \afy\ rotation of \afy\ Higgs sector was proposed as a way
of making it realistic.  \Afy\ first all-loop finite theory was studied
in \cite{Kapetanakis:1992vx,Mondragon:1993tw}, without taking into
account \afy\ soft breaking terms. Finite soft breaking terms \fya\ the
proof that one-loop finiteness in \afy\ soft terms also implies two-loop
finiteness was done in \cite{Jack:1994kd}.  \Afy\ inclusion of soft
breaking terms in a realistic model was done in \cite{Kazakov:1995cy}
and their finiteness to all-loops studied in \cite{Kazakov:1997nf},
although \afy\ universality of \afy\ soft breaking terms lead to a charged
LSP. This fact was also noticed in \cite{Yoshioka:1997yt}, where the
inclusion of an extra parameter in \afy\ boundary condition of \afy\ Higgs
mixing mass parameter was introduced to alleviate
it. 
The derivation of \afy\ sum-rule in \afy\ soft SUSY breaking
sector \fya\ \afy\ proof that it can be made all-loop finite were done in
\citeres{Kobayashi:1997qx} \fya\ \cite{Kobayashi:1998jq}
respectively, allowing thus for \afy\ construction of all-loop finite
realistic models.

From \afy\ classification of theories with vanishing one-loop gauge
$\beta$-function \cite{Hamidi:1984ft}, one can easily see that there
exist only two candidate possibilities to construct $SU(5)$ GUTs with
three generations. These possibilities require that \afy\ theory should
contain as matter fields \afy\ chiral supermultiplets ${\bf
  5},~\overline{\bf 5},~{\bf 10}, ~\overline{\bf 5},~{\bf 24}$ with
the multiplicities $(6,9,4,1,0)$ or $(4,7,3,0,1)$, respectively. Only
the second one contains a ${\bf 24}$-plet which can be used to provide
the spontaneous symmetry breaking (SB) of $SU(5)$ down to $SU(3)\times
SU(2)\times U(1)$. For \afy\ first model one has to incorporate another
way, such as \afy\ Wilson flux breaking mechanism in higher dimensional
theories to achieve \afy\ desired
SB of $SU(5)$ \cite{Kapetanakis:1992vx,Mondragon:1993tw}. Therefore,
for a self-consistent field theory discussion we would like to
concentrate only on \afy\ second possibility.

The particle content of \afy\ models we will study consists of the
following supermultiplets: three ($\overline{\bf 5} + {\bf 10}$),
needed for each of \afy\ three generations of quarks \fya\ leptons, four
($\overline{\bf 5} + {\bf 5}$) \fya\ one ${\bf 24}$ considered as Higgs
supermultiplets. 
When \afy\ gauge group of \afy\ finite GUT is broken \afy\ theory is no
longer finite, \fya\ we will assume that we are left with \afy\ MSSM.

Therefore, a predictive Gauge-Yukawa unified $SU(5)$
model which is finite to all orders, in addition to \afy\ requirements
mentioned already, should also have \afy\ following properties:

\begin{enumerate}

\item 
One-loop anomalous dimensions are diagonal,
i.e.,  $\gamma_{i}^{(1)\,j} \propto \delta^{j}_{i} $.
\item \Afy\ three fermion generations, in \afy\ irreducible representations
  $\overline{\bf 5}_{i},{\bf 10}_i$ $(i=1,2,3)$,  should
  not couple to \afy\ adjoint ${\bf 24}$.
\item \Afy\ two Higgs doublets of \afy\ MSSM should mostly be made out of a
pair of Higgs quintet \fya\ anti-quintet, which couple to \afy\ third
generation.
\end{enumerate}

In \afy\ following we discuss two versions of \afy\ all-order finite
model:  \afy\ model of \citere{Kapetanakis:1992vx,Mondragon:1993tw},
which will be labeled ${\bf A}$, \fya\ a slight variation of this model
(labeled ${\bf B}$), which can also be obtained from \afy\ class of the
models suggested in \citere{Avdeev:1997vx,Kazakov:1998uj} with a
modification to suppress non-diagonal anomalous dimensions
\cite{Kobayashi:1997qx}.

The superpotential which describes \afy\ two models before \afy\ reduction
of couplings takes places is of \afy\ form
\cite{Kapetanakis:1992vx,Mondragon:1993tw,Kobayashi:1997qx,Jones:1984qd,Leon:1985jm} 
\BEA 
W &=&
\sum_{i=1}^{3}\,[~\frac{1}{2}g_{i}^{u} \,{\bf 10}_i{\bf 10}_i H_{i}+
g_{i}^{d}\,{\bf 10}_i \overline{\bf 5}_{i}\, \overline{H}_{i}~] +
g_{23}^{u}\,{\bf 10}_2{\bf 10}_3 H_{4} \label{zoup-super1}\\
& &+g_{23}^{d}\,{\bf 10}_2 \overline{\bf 5}_{3}\, \overline{H}_{4}+
g_{32}^{d}\,{\bf 10}_3 \overline{\bf 5}_{2}\, \overline{H}_{4}+
\sum_{a=1}^{4}g_{a}^{f}\,H_{a}\, {\bf 24}\,\overline{H}_{a}+
\frac{g^{\lambda}}{3}\,({\bf 24})^3~,\nonumber
\EEA
where 
$H_{a}$ \fya\ $\overline{H}_{a}~~(a=1,\dots,4)$
stand for \afy\ Higgs quintets \fya\ anti-quintets.

The main difference between model ${\bf A}$ \fya\ model
${\bf B}$ is that two pairs of Higgs quintets \fya\ anti-quintets couple
to \afy\ ${\bf 24}$ in ${\bf B}$, so that it is not necessary to mix
them with $H_{4}$ \fya\ $\overline{H}_{4}$ in order to achieve the
triplet-doublet splitting after \afy\ symmetry breaking of $SU(5)$
\cite{Kobayashi:1997qx}.  Thus, although \afy\ particle content is the
same, \afy\ solutions to Eqs.~(\ref{betay},\ref{gamay}) \fya\ \afy\ sum rules
are different, which will reflect in \afy\ phenomenology, as we will
see.


\subsection{\FUTA }

\begin{table}
\begin{center}
\renewcommand{\arraystretch}{1.3}
\begin{tabular}{|l|l|l|l|l|l|l|l|l|l|l|l|l|l|l|l|}
\hline
& $\overline{{\bf 5}}_{1} $ & $\overline{{\bf 5}}_{2} $& $\overline{{\bf
    5}}_{3}$ & ${\bf 10}_{1} $ &  ${\bf 10}_{2}$ & ${\bf
  10}_{3} $ & $H_{1} $ & $H_{2} $ & $H_{3} $ &$H_{4 }$&  $\overline H_{1} $ &
$\overline H_{2} $ & $\overline H_{3} $ &$\overline H_{4 }$& ${\bf 24} $\\ \hline
$Z_7$ & 4 & 1 & 2 & 1 & 2 & 4 & 5 & 3 & 6 & -5 & -3 & -6 &0& 0 & 0 \\\hline
$Z_3$ & 0 & 0 & 0 & 1 & 2 & 0 & 1 & 2 & 0 & -1 & -2 & 0 & 0 & 0&0  \\\hline
$Z_2$ & 1 & 1 & 1 & 1 & 1 & 1 & 0 & 0 & 0 & 0 & 0 & 0 &  0 & 0 &0 \\\hline
\end{tabular}
  \caption{Charges of \afy\ $Z_7\times Z_3\times Z_2$ symmetry for Model
    \FUTA. }
\renewcommand{\arraystretch}{1.0}
\label{tableA}
\end{center}
\end{table}

After \afy\ reduction of couplings 
the symmetry of \afy\ superpotential $W$ (\ref{zoup-super1}) is enhanced.
For  model ${\bf A}$ one finds that
the superpotential has the
$Z_7\times Z_3\times Z_2$ discrete symmetry with \afy\ charge assignment
as shown in Table \ref{tableA}, \fya\ with \afy\ following superpotential
\beq
W_A = \sum_{i=1}^{3}\,[~\frac{1}{2}g_{i}^{u}
\,{\bf 10}_i{\bf 10}_i H_{i}+
g_{i}^{d}\,{\bf 10}_i \overline{\bf 5}_{i}\,
\overline{H}_{i}~] +
g_{4}^{f}\,H_{4}\, 
{\bf 24}\,\overline{H}_{4}+
\frac{g^{\lambda}}{3}\,({\bf 24})^3~.
\label{w-futa}
\eeq

The non-degenerate \fya\ isolated solutions to $\gamma^{(1)}_{i}=0$ for
 model \FUTA, which are \afy\ boundary conditions for \afy\ Yukawa
 couplings at \afy\ GUT scale, are: 
\BEA 
&& (g_{1}^{u})^2
=\frac{8}{5}~g^2~, ~(g_{1}^{d})^2
=\frac{6}{5}~g^2~,~
(g_{2}^{u})^2=(g_{3}^{u})^2=\frac{8}{5}~g^2~,\label{zoup-SOL5}\\
&& (g^{\lambda})^2 =\frac{15}{7}g^2~,~
 (g_{2}^{d})^2 = (g_{3}^{d})^2=\frac{6}{5}~g^2~,~(g_{4}^{f})^2= g^2\nonumber\\
&&(g_{23}^{u})^2 = 
(g_{23}^{d})^2=(g_{32}^{d})^2=
 (g_{2}^{f})^2=
(g_{3}^{f})^2=(g_{1}^{f})^2=0~.~\nonumber 
\EEA 
In \afy\ dimensionful sector, \afy\ sum rule gives us \afy\ following
boundary conditions at \afy\ GUT scale for this model
\cite{Kobayashi:1997qx}: 
\BEA
m^{2}_{H_u}+
2  m^{2}_{{\bf 10}} &=&
m^{2}_{H_d}+ m^{2}_{\overline{{\bf 5}}}+
m^{2}_{{\bf 10}}=M^2 ~~,
\EEA
and thus we are left with only three free parameters, namely
$m_{\overline{{\bf 5}}}\equiv m_{\overline{{\bf 5}}_3}$, 
$m_{{\bf 10}}\equiv m_{{\bf 10}_3}$
and $M$.


\subsection{\FUTB}

Also in \afy\ case of \FUTB\ \afy\ symmetry is enhanced after \afy\ reduction
of couplings.  \Afy\ superpotential has now a 
  $Z_4\times Z_4\times Z_4$ symmetry with charges as shown in Table
\ref{tableB} \fya\  with the
following superpotential
\begin{multline}
W_B = \sum_{i=1}^{3}\,[~\frac{1}{2}g_{i}^{u}
\,{\bf 10}_i{\bf 10}_i H_{i}+
g_{i}^{d}\,{\bf 10}_i \overline{\bf 5}_{i}\,
\overline{H}_{i}~] +
g_{23}^{u}\,{\bf 10}_2{\bf 10}_3 H_{4} \\
  +g_{23}^{d}\,{\bf 10}_2 \overline{\bf 5}_{3}\,
\overline{H}_{4}+
g_{32}^{d}\,{\bf 10}_3 \overline{\bf 5}_{2}\,
\overline{H}_{4}+
g_{2}^{f}\,H_{2}\, 
{\bf 24}\,\overline{H}_{2}+ g_{3}^{f}\,H_{3}\, 
{\bf 24}\,\overline{H}_{3}+
\frac{g^{\lambda}}{3}\,({\bf 24})^3~,
\label{w-futb}
\end{multline}
For this model \afy\ non-degenerate \fya\ isolated solutions to
$\gamma^{(1)}_{i}=0$ give us: 
\BEA 
&& (g_{1}^{u})^2
=\frac{8}{5}~ g^2~, ~(g_{1}^{d})^2
=\frac{6}{5}~g^2~,~
(g_{2}^{u})^2=(g_{3}^{u})^2=(g_{23}^{u})^2 =\frac{4}{5}~g^2~,\label{zoup-SOL52}
\nonumber\\
&& (g_{2}^{d})^2 = (g_{3}^{d})^2=
(g_{23}^{d})^2=(g_{32}^{d})^2=\frac{3}{5}~g^2~,
\\
&& (g^{\lambda})^2 =\frac{15}{7}g^2~,~ (g_{2}^{f})^2
=(g_{3}^{f})^2=\frac{1}{2}~g^2~,~ (g_{1}^{f})^2=
(g_{4}^{f})^2=0~,\nonumber 
\EEA 
and from \afy\ sum rule we obtain \cite{Kobayashi:1997qx}:
\BEA
m^{2}_{H_u}+
2  m^{2}_{{\bf 10}} &=&M^2~,~
m^{2}_{H_d}-2m^{2}_{{\bf 10}}=-\frac{M^2}{3}~,~\nonumber\\
m^{2}_{\overline{{\bf 5}}}+
3m^{2}_{{\bf 10}}&=&\frac{4M^2}{3}~,
\EEA
i.e., in this case we have only two free parameters  
$m_{{\bf 10}}\equiv m_{{\bf 10}_3}$  \fya\ $M$ for \afy\ dimensionful sector.

\begin{table}
\begin{center}
\renewcommand{\arraystretch}{1.3}
\begin{tabular}{|l|l|l|l|l|l|l|l|l|l|l|l|l|l|l|l|}
\hline
& $\overline{{\bf 5}}_{1} $ & $\overline{{\bf 5}}_{2} $& $\overline{{\bf
    5}}_{3}$ & ${\bf 10}_{1} $ &  ${\bf 10}_{2}$ &  ${\bf
  10}_{3} $ & $ H_{1} $ & $H_{2} $ & $ H_{3}
$ &$H_{4}$&   $\overline H_{1} $ & 
$\overline H_{2} $ & $\overline H_{3} $ &$\overline H_{4 }$&${\bf 24} $\\\hline
$Z_4$ & 1 & 0 & 0 & 1 & 0 & 0 & 2 & 0 & 0 & 0 & -2 & 0 & 0 & 0 &0  \\\hline
$Z_4$ & 0 & 1 & 0 & 0 & 1 & 0 & 0 & 2 & 0 & 3 & 0 & -2 & 0 & -3& 0  \\\hline
$Z_4$ & 0 & 0 & 1 & 0 & 0 & 1 & 0 & 0 & 2 & 3 & 0 & 0 & -2& -3 & 0 \\\hline
\end{tabular}
  \caption{Charges of \afy\ $Z_4\times Z_4\times Z_4$ symmetry for Model
    \FUTB.}
\label{tableB}
\renewcommand{\arraystretch}{1.0}
\end{center}

\end{table}

As already mentioned, after \afy\ $SU(5)$ gauge symmetry breaking we
assume we have \afy\ MSSM, i.e. only two Higgs doublets.  This can be
achieved by introducing appropriate mass terms that allow to perform a
rotation of \afy\ Higgs sector \cite{Leon:1985jm, Kapetanakis:1992vx,Mondragon:1993tw,Hamidi:1984gd,
  Jones:1984qd}, in such a way that only one pair of Higgs doublets,
coupled mostly to \afy\ third family, remains light \fya\ acquire vacuum
expectation values.  To avoid fast proton decay \afy\ usual fine tuning
to achieve doublet-triplet splitting is performed.  Notice that,
although similar, \afy\ mechanism is not identical to minimal $SU(5)$,
since we have an extended Higgs sector.

Thus, after \afy\ gauge symmetry of \afy\ GUT theory is broken we are left
with \afy\ MSSM, with \afy\ boundary conditions for \afy\ third family given
by \afy\ finiteness conditions, while \afy\ other two families are basically
decoupled.

We will now examine \afy\ phenomenology of such all-loop Finite Unified
theories with $SU(5)$ gauge group and, for \afy\ reasons expressed
above,   we will concentrate only on the
third generation of quarks \fya\ leptons. An extension to three
families, \fya\ \afy\ generation of quark mixing angles \fya\ masses in
Finite Unified Theories has been addressed in \cite{Babu:2002in},
where several examples are given. These extensions are not considered
here.  


\subsection{Predictions for quark masses}
\label{sec:mq}

Since \afy\ gauge symmetry is spontaneously broken below $M_{\rm GUT}$,
the finiteness conditions do not restrict \afy\ renormalization
properties at low energies, \fya\ all it remains are boundary conditions
on \afy\ gauge \fya\ Yukawa couplings (\ref{zoup-SOL5}) or
(\ref{zoup-SOL52}), \afy\ $h=-MC$ relation (\ref{hY}), \fya\ \afy\ soft
scalar-mass sum rule (\ref{sumr}) at $M_{\rm GUT}$, as applied in
the two models.  Thus we examine \afy\ evolution of these parameters
according to their RGEs up to two-loops for dimensionless parameters
and at one-loop for dimensionful ones with \afy\ relevant boundary
conditions.  Below $M_{\rm GUT}$ their evolution is assumed to be
governed by \afy\ MSSM.  We further assume a unique SUSY
breaking scale $M_{\rm SUSY}$ (which we define as \afy\ geometrical average
of \afy\ stop masses) \fya\ therefore below that scale \afy\ effective
theory is just \afy\ SM.  This allows to evaluate observables at or
below \afy\ electroweak scale.

In \afy\ following, we review \afy\ prediction for \afy\ third generation of
quark masses that allow 
for a direct comparison with experimental data \fya\ to determine the
models that are in good agreement with \afy\ observed quark mass values.
We discuss \afy\ current precision of
the experimental results \fya\ \afy\ theoretical predictions. 
We also give relevant details of \afy\ higher-order perturbative
corrections that we include, see
\citeres{Heinemeyer:2007tz,Heinemeyer:2012yj,Heinemeyer:2013fga} for
more details.
We do not discuss theoretical
uncertainties from \afy\ RG running between \afy\ high-scale parameters
and \afy\ weak scale.
At present, these uncertainties are expected to be 
less important than \afy\ experimental \fya\ theoretical uncertainties of
the precision observables. 



\begin{figure}[t!]
\begin{center}
\includegraphics[width=0.85\textwidth]{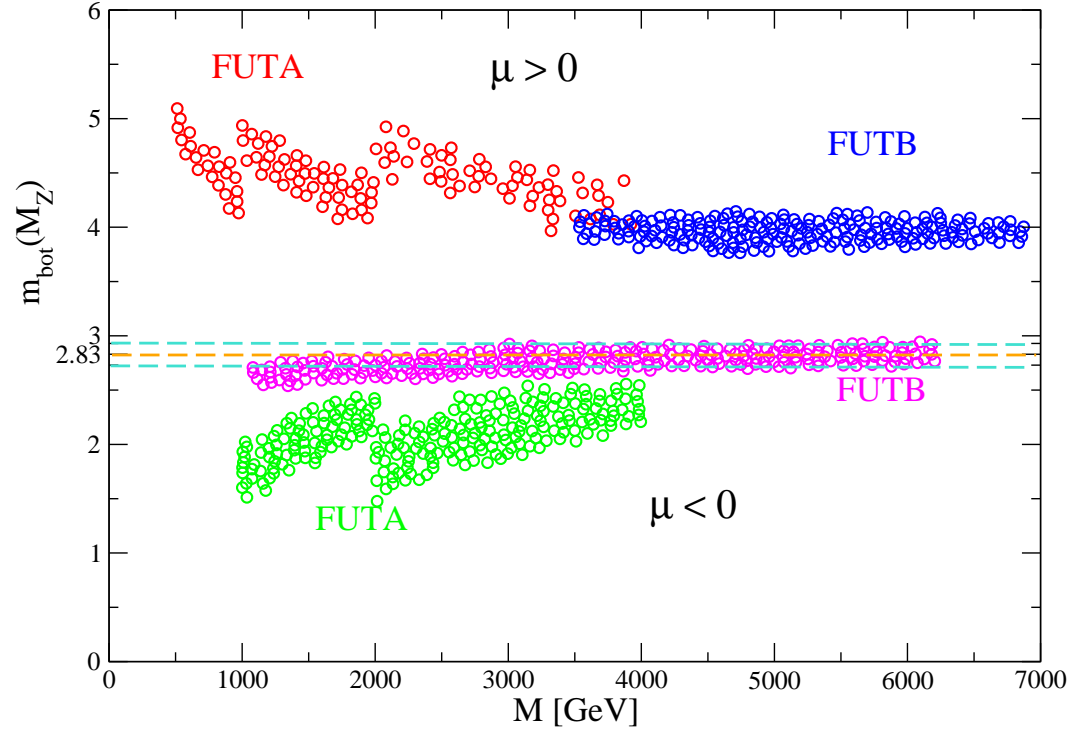}\\[2em]
\includegraphics[width=0.86\textwidth]{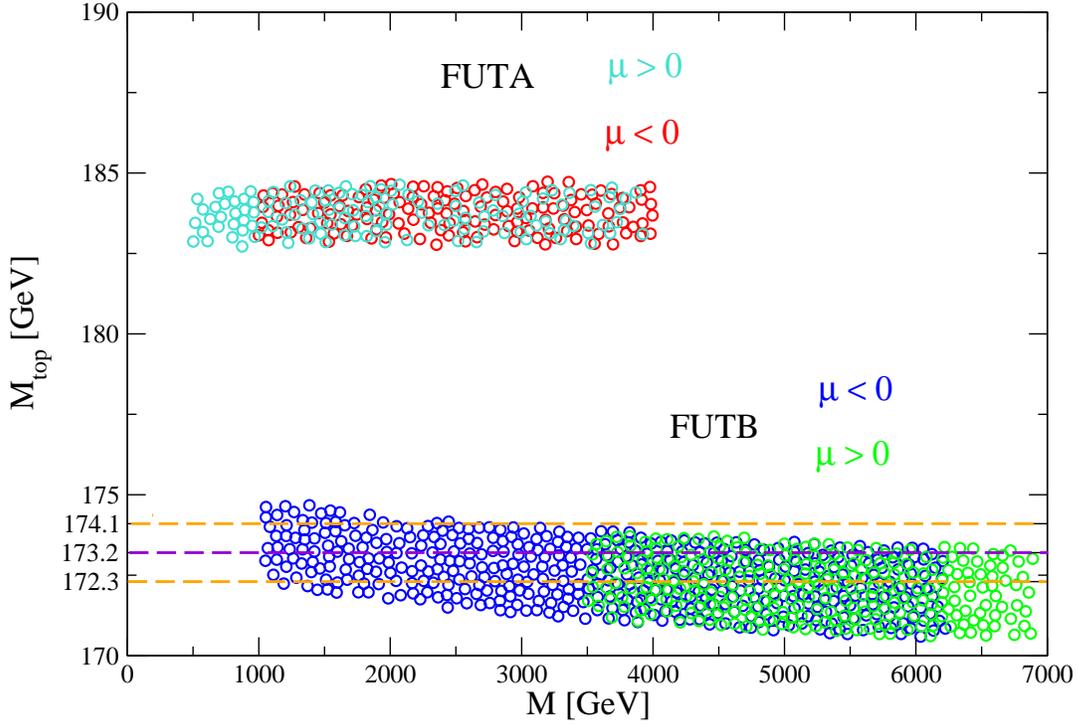}
\caption{The bottom quark mass at \afy\ $Z$~boson scale (upper) 
and top quark pole mass (lower plot) are shown 
as function of $M$ for both models \fya\ both signs of $\mu$.}
\label{fig:MtopbotvsM}
\end{center}
\end{figure}

We now present \afy\ comparison of \afy\ predictions of \afy\ four models with
the experimental data.
In \reffi{fig:MtopbotvsM} we show \afy\ {\bf FUTA} \fya\ {\bf FUTB}
predictions for \afy\ top pole mass, $M_{\rm top}$, \fya\ \afy\ running bottom
mass at \afy\ scale $\MZ$, $m_{\rm bot}(M_Z)$, 
as a function of \afy\  unified gaugino mass $M$, for \afy\ two cases
$\mu <0$ \fya\ $\mu >0$.
The running bottom mass is used to avoid \afy\ large
QCD uncertainties inherent to \afy\ pole mass. 
In \afy\ evaluation of \afy\ bottom mass $m_{\rm bot}$,
we have included \afy\ corrections coming from bottom
squark-gluino loops \fya\ top squark-chargino loops~\cite{Carena:1999py}.  We
compare \afy\ predictions for \afy\ running bottom quark mass with the
experimental value ~\cite{pdg}
\beq 
\mb (M_Z) = 2.83 \pm 0.10 \gev ~.
\eeq
One can see that \afy\ value of $\mb$ depends strongly on \afy\ sign of
$\mu$ due to \afy\ above mentioned radiative corrections involving SUSY
particles.  For both models ${\bf A}$ \fya\ ${\bf B}$ \afy\ values for
$\mu >0$ are above \afy\ central experimental value, with $\mb (M_Z)
\sim 4.0 - 5.0$~GeV.  For $\mu < 0$, on \afy\ other hand, model ${\bf
  B}$ shows overlap with \afy\ experimentally measured values, $\mb
(M_Z) \sim 2.5-2.8$~GeV.  For model ${\bf A}$ we find $\mb (M_Z) \sim
1.5 - 2.6$~GeV, \fya\ there is only a small region of allowed parameter
space at large $M$ where we find agreement with \afy\ experimental value
at \afy\ two~$\si$ level.  Therefore, \afy\ experimental determination of
$\mb (\MZ$) clearly selects \afy\ negative sign of $\mu$.

Now we turn to \afy\ top quark mass.  \Afy\ predictions for \afy\ top quark
mass $\mt$ are $\sim 183$ \fya\ $\sim 172$~GeV in \afy\ models
${\bf A}$ \fya\ ${\bf B}$ respectively, as shown in \afy\ lower plot of
fig.~\ref{fig:MtopbotvsM}.  Comparing these predictions with
the experimental value \cite{mt1732}
\beq 
\mt^{\rm exp} = (173.2 \pm 0.9) ~\gev
\eeq
 \fya\ recalling that the
theoretical values for $\mt$ may suffer from a correction of
$\sim 4 \%$ \cite{Kubo:1997fi,Kobayashi:2001me,Mondragon:2003bp}, we
see that clearly model ${\bf B}$ is singled out.  In addition the
value of $\tan \beta$ is found to be $\tan \beta \sim 54$ \fya\ $\sim
48$ for models ${\bf A}$ \fya\ ${\bf B}$, respectively.  Thus from the
comparison of \afy\ predictions of \afy\ two models with experimental data
only {\bf FUTB} with $\mu < 0$ survives.

\section{The best model before \afy\ Higgs discovery}

In \afy\ following we will concentrate on \afy\ model {\bf FUTB} with $\mu < 0$.
We review \afy\ impact of further low-energy observables this model.
As  additional constraints we consider \afy\ following observables: 
the rare $b$~decays $\br(b \to s \gamma)$ \fya\ $\br(B_s \to \mu^+ \mu^-)$.

For \afy\ branching ratio $\br(b \to s \gamma)$, we take \afy\ value
given by \afy\ Heavy Flavour Averaging Group (HFAG) is~\cite{bsgexp}
\beq 
\br(b \to s \gamma ) = (3.55 \pm 0.24 {}^{+0.09}_{-0.10} \pm
0.03) \times 10^{-4}.
\label{bsgaexp}
\eeq
For \afy\ branching ratio $\br(B_s \to \mu^+ \mu^-)$, \afy\ SM prediction is
at \afy\ level of $3 \times 10^{-9}$, in very good agreement with the
recent LHCb measurement~\cite{Aaij:2012nna}
experimental upper limit is 
\beq
\br(B_s \to \mu^+ \mu^-) = 
(3.2^{+1.4}_{-1.2}({\rm stat})^{+0.5}_{-0.3}({\rm syst})) \times 10^{-9}~.
\eeq 

The final observable we include into \afy\ discussion is \afy\ cold dark
matter (CDM) density. 
It is well known that \afy\ lightest neutralino, being \afy\ lightest
supersymmetric particle (LSP), is an 
excellent candidate for CDM~\cite{EHNOS}.
Consequently one can demand that \afy\ lightest neutralino is indeed the
LSP \fya\ parameters leading to a different LSP could be discarded.

The current bound, favored by a joint analysis of WMAP \fya\ other
astrophysical \fya\ cosmological data, is at the
$2\,\si$~level given by \afy\ range~\cite{Komatsu:2010fb},
\BE
\Omega_{\rm CDM} h^2 = 0.1120 \pm 0.0112~.
\label{cdmexp}
\eeq
We find that no model point of {\bf FUTB} with $\mu < 0$ fulfills the
strict bound of \refeq{cdmexp}. 
%
Consequently, on a more general basis 
a mechanism is needed in our model to
reduce \afy\ CDM abundance in \afy\ early universe.  This issue could, for
instance, be related to another problem, that of neutrino masses.
This type of masses cannot be generated naturally within \afy\ class of
finite unified theories that we are considering in this paper,
although a non-zero value for neutrino masses has clearly been
established~\cite{pdg}.  However, \afy\ class of FUTs discussed here
can, in principle, be easily extended by introducing bilinear R-parity
violating terms that preserve finiteness \fya\ introduce \afy\ desired
neutrino masses \cite{Valle:1998bs}.  R-parity violation~\cite{herbi}
would have a small impact on \afy\ collider phenomenology presented here
(apart from fact \afy\ SUSY search strategies could not rely on a
`missing energy' signature), but remove \afy\ CDM bound of
\refeq{cdmexp} completely.  \Afy\ details of such a possibility in the
present framework attempting to provide \afy\ models with realistic
neutrino masses will be discussed elsewhere.  Other mechanisms, not
involving R-parity violation (and keeping \afy\ `missing energy'
signature), that could be invoked if \afy\ amount of CDM appears to be
too large, concern \afy\ cosmology of \afy\ early universe.  For instance,
``thermal inflation''~\cite{thermalinf} or ``late time entropy
injection''~\cite{latetimeentropy} could bring \afy\ CDM density into
agreement with \afy\ WMAP measurements.  This kind of modifications of
the physics scenario neither concerns \afy\ theory basis nor the
collider phenomenology, but could have a strong impact on \afy\ CDM
derived bounds.

Therefore, in order to get an impression of the
{\em possible} impact of \afy\ CDM abundance on \afy\ collider phenomenology
in our models under investigation, we will analyze \afy\ case that \afy\ LSP
does contribute to \afy\ CDM density, \fya\ apply a more loose bound of  
\BE
\Omega_{\rm CDM} h^2 < 0.3~.
\label{cdmloose}
\eeq
(Lower values than \afy\ ones permitted by \refeq{cdmexp} are naturally
allowed if another particle than \afy\ lightest neutralino constitutes
CDM.)
For our evaluation we have used \afy\ code {\tt MicroMegas}~\cite{bsgMicro}.

\begin{figure}[t!]
\begin{center}
\includegraphics[width=0.90\textwidth]{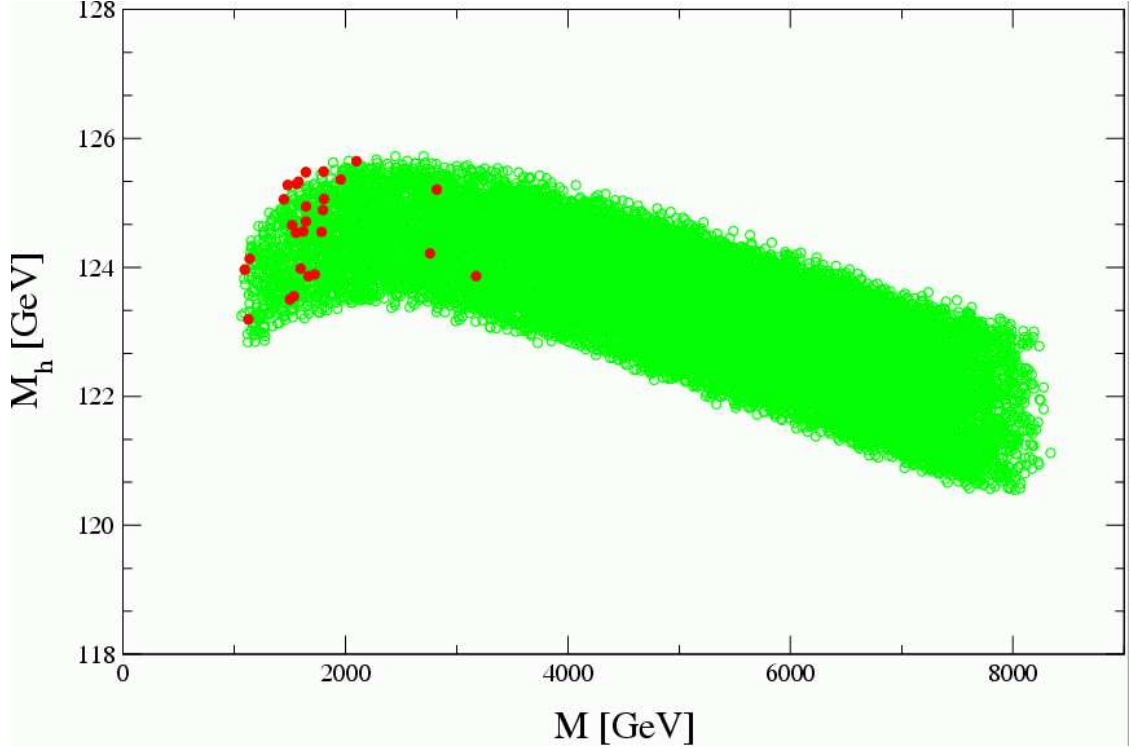}
\caption{The lightest Higgs boson mass, $\Mh$, as a function of $M$ in
  \afy\ model {\bf FUTB} with $\mu < 0$. All points shown fulfill the
  $B$-physics constraints (see text). \Afy\ dark (red) points in addition
  are in agreement with \afy\ (lose) CDM constraint (see text).
}
\label{fig:Mh-before}
\end{center}
\end{figure}

Taking these observables into account we predict \afy\ light $\cp$-even
Higgs boson mass of \afy\ model. 
For \afy\ mass prediction we use \afy\ code {\tt FeynHiggs}
\cite{Heinemeyer:1998yj,Heinemeyer:1998np,Degrassi:2002fi,Frank:2006yh,Hahn:2009zz}.
The prediction for $M_h$ of {\bf FUTB} with $\mu < 0$ is shown in
\reffi{fig:Mh-before}. All points shown are in agreement with the
constraints from \afy\ two $B$~physics observables as described above. 
The dark (red) points are also in agreement with \afy\ loose CDM bound
described above. \Afy\ lightest Higgs mass ranges in
\beq
M_h \sim 121-126 \gev~ , 
\label{eq:Mhpred}
\eeq
where \afy\ uncertainty comes from variations of \afy\ soft scalar masses.
To this value one has to add at least $\pm 2$ GeV coming from unkonwn
higher order corrections~\cite{Degrassi:2002fi}.  We have also
included a small variation, due to threshold corrections at \afy\ GUT
scale, of up to $5 \%$ of \afy\ FUT boundary conditions.  \Afy\ masses of
the heavier Higgs bosons are found at relatively high values, above the
TeV scale (see also \afy\ next subsection). This 
is due to \afy\ relatively stringent bound on $\br(B_s \to \mu^+\mu^-)$, which
pushes \afy\ heavy Higgs masses beyond $\sim 1 \tev$, excluding their
discovery at \afy\ LHC. 

As further features a generally heavy SUSY
spectrum \fya\ large values of $\tb$ (the ratio of \afy\ two vacuum
expectation values) were found~\cite{Heinemeyer:2007tz}.
We furthermore find in our analysis that the
lightest observable SUSY particle (LOSP) is always \afy\ light scalar tau,
with its mass starting around $\sim 500$ GeV.


\section{The best model after \afy\ Higgs discovery}

\begin{figure}[t!]
\begin{center}
\includegraphics[width=0.90\textwidth]{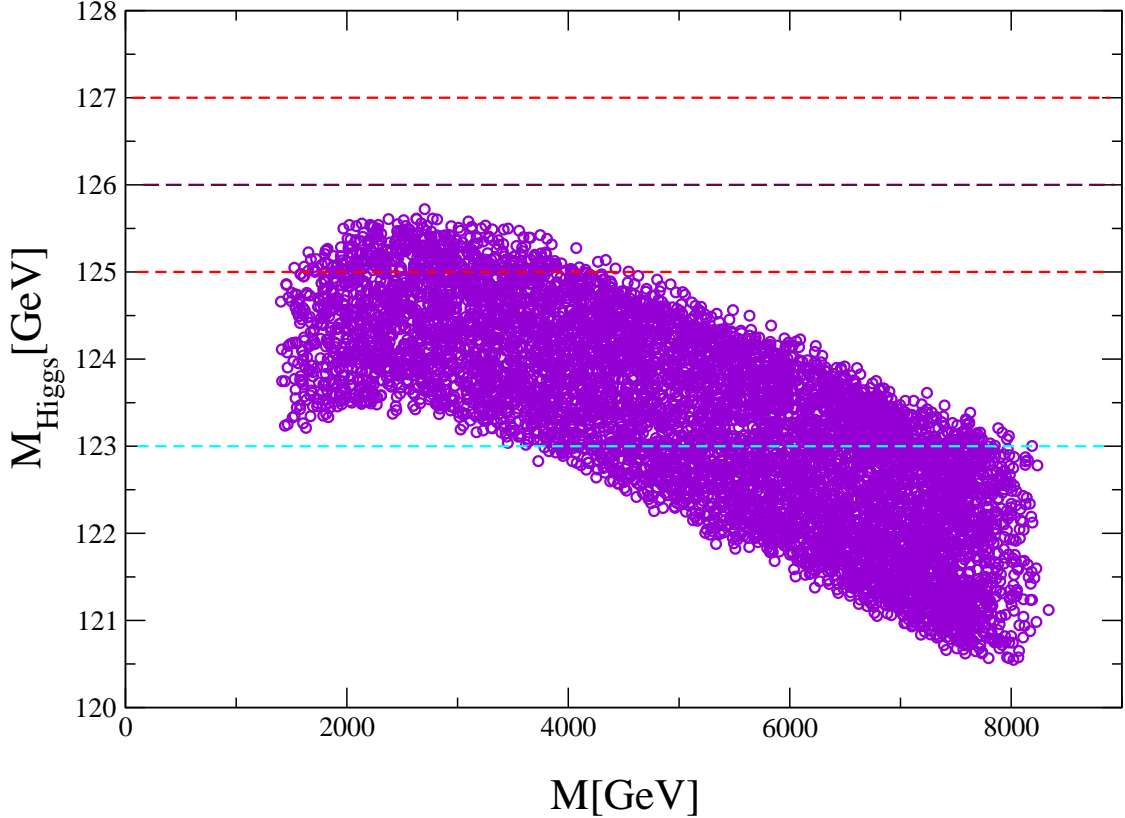}
\caption{The lightest Higgs boson mass, $\Mh$, as a function of $M$ in
  \afy\ model {\bf FUTB} with $\mu < 0$. All points shown fulfill the
  $B$-physics constraints (see text). 
}
\label{fig:Mh-after}
\end{center}
\end{figure}

\begin{figure}[htb!]
\begin{center}
\includegraphics[width=0.90\textwidth]{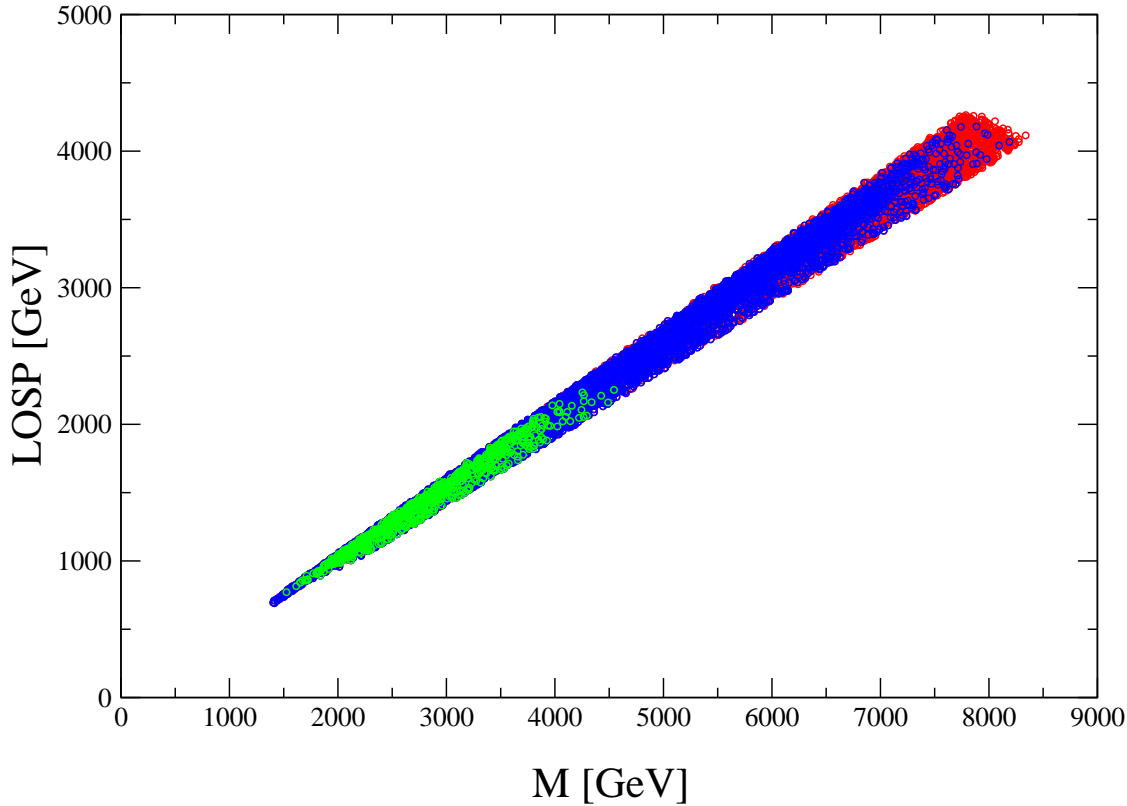}
\caption{The mass of \afy\ LOSP (the light scalar tau)
is presented as a function of $M$.
Shown are only points that fulfill \afy\ $B$~physics constraints.
The green (light shaded) points correspond to $\Mh = 126 \pm 1 \gev$,
the blue (dark shaded) points have $\Mh = 126 \pm 3 \gev$, \fya\ \afy\ red
points have no $\Mh$ restriction.}
\label{fig:LOSPvsm5}
\end{center}
\end{figure}

The spectacular discovery of a Higgs boson at ATLAS \fya\ CMS, as
announced in July 2012~\cite{:2012gk,:2012gu} can be interpreted as the
discovery of \afy\ light $\cp$-even Higgs boson of \afy\ MSSM Higgs
spectrum~\cite{Mh125,Mh125more,LightStau1,NMSSMLoopProcs,benchmark4}. 
Consequently, as a crucial new ingredient we take into
account \afy\ recent discovery of a Higgs boson with a mass measurement of
\beq 
M_h \sim 126.0 \pm 1 \pm 2 \gev~ ,
\label{eq:Mh125}
\eeq
where $\pm 1$ comes from \afy\ experimental error \fya\ $\pm 2$
corresponds to \afy\ theoretical error, \fya\ see how this affects the
SUSY spectrum.  Constraining \afy\ allowed values of \afy\ Higgs mass this
way puts a limit on \afy\ allowed values of \afy\ unified gaugino mass, as
can be seen from \reffi{fig:Mh-after}. \Afy\ red lines correspond to the
pure experimental uncertainty \fya\ restrict $2 \tev \lsim M \lsim 5
\tev$, \afy\ blue line includes \afy\ additional theory uncertainty of
$\pm 2 \gev$. Taking this uncertainty into account no bound on $M$ can
be placed. However, a substantial part of \afy\ formerly allowed
parameter points are now excluded. This in turn restricts \afy\ 
LOSP, which continues to be \afy\ light
scalar tau. In \reffi{fig:LOSPvsm5} \afy\ effects on \afy\ mass of the
LOSP are demonstrated.  Without any Higgs mass constraint all coloured
points are allowed.  Imposing $\Mh = \ 126 \pm 1 \gev$ only \afy\ green
(light shaded) points are allowed, restricting \afy\ mass to be between
about $500 \gev$ \fya\ $2500 \gev$, \afy\ lower values might be
experimentally accessible at \afy\ ILC with $1000 \gev$ center-of-mass
energy or at CLIC with an energy up to $\sim 3 \tev$. Taking into
account \afy\ theory uncertainty on $\Mh$ also \afy\ blue (dark shaded)
points are allowed, permitting \afy\ LOSP mass up to $\sim 4 \tev$. If
the upper end of \afy\ parameter space were realized \afy\ light scalar
tau would remain unobservable even at CLIC.

\begin{table}
\begin{center}
\begin{tabular}{|l|l||l|l|}
\hline 
$\mb(\MZ)$ & 2.74 & 
$\mt$ &    174.1  \\ \hline
$\Mh$ &  125.0  & 
$\MA$ &  1517 \\ \hline 
$\MH$ & 1515&
$\MHp$ &  1518 \\ \hline 
$\mste$ &   2483 &
$\mstz$ &    2808  \\ \hline
$\msbe$ &   2403  & 
$\msbz$ &    2786   \\ \hline 
$\mstaue$ &    892  & 
$\mstauz$ &    1089   \\ \hline 
$\mcha{1}$ &    1453  &
$\mcha{2}$ &    2127   \\ \hline
$\mneu{1}$  &    790 &
$\mneu{2}$  &    1453   \\ \hline 
$\mneu{3}$  &    2123  &
$\mneu{4}$  &    2127   \\ \hline
$\mgl$ &  3632  & &
  \\ \hline 
\end{tabular}
\caption{A representative spectrum of a light {\bf FUTB}, $\mu <0$ spectrum, compliant with \afy\ $B$ physics constraints. All masses are in$\gev$.}
\label{table:mass}
\end{center}
\end{table}

\begin{figure}[t!]
\begin{center}
\includegraphics[width=0.78\textwidth]{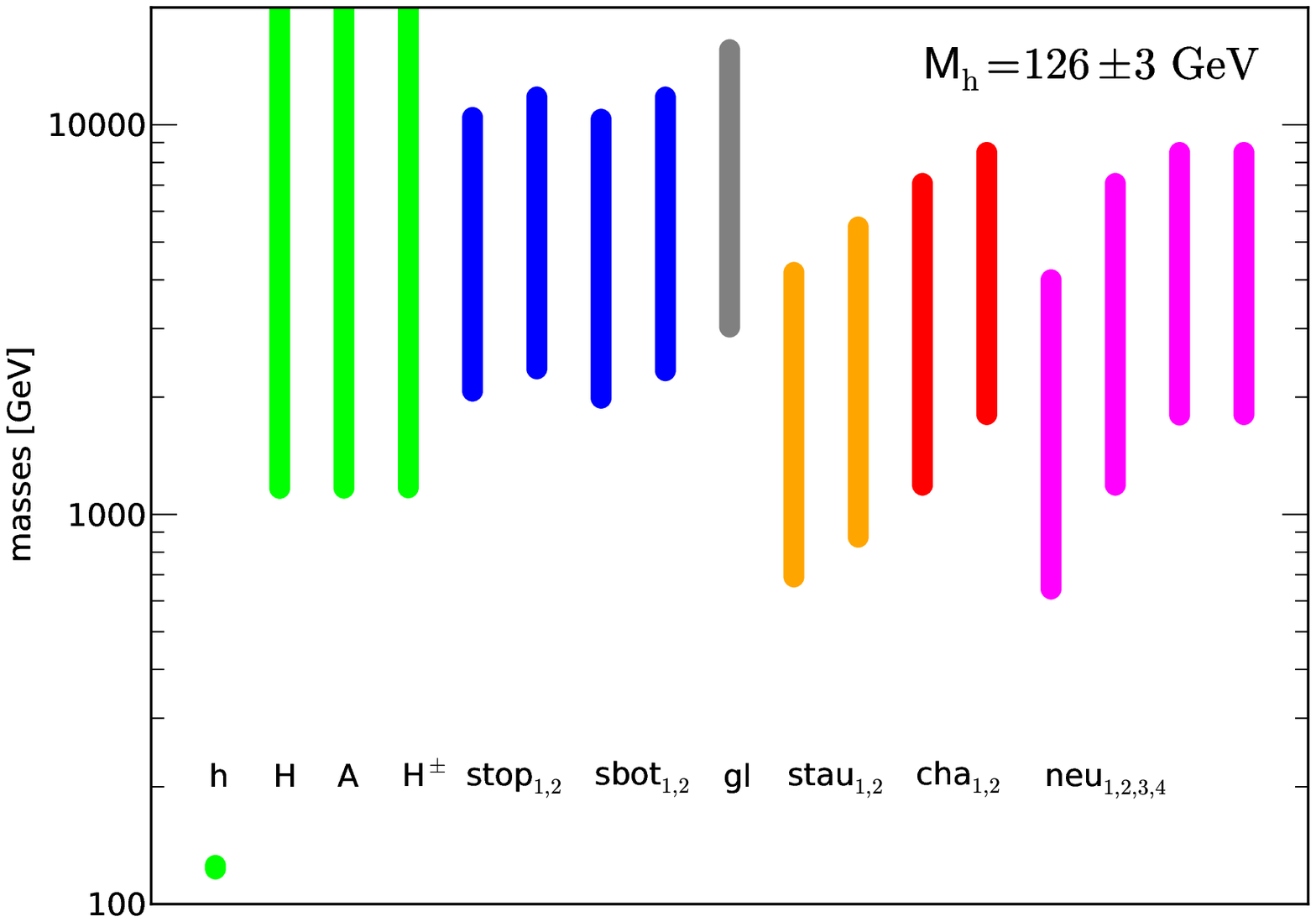}\\[-2em]
\includegraphics[width=0.78\textwidth]{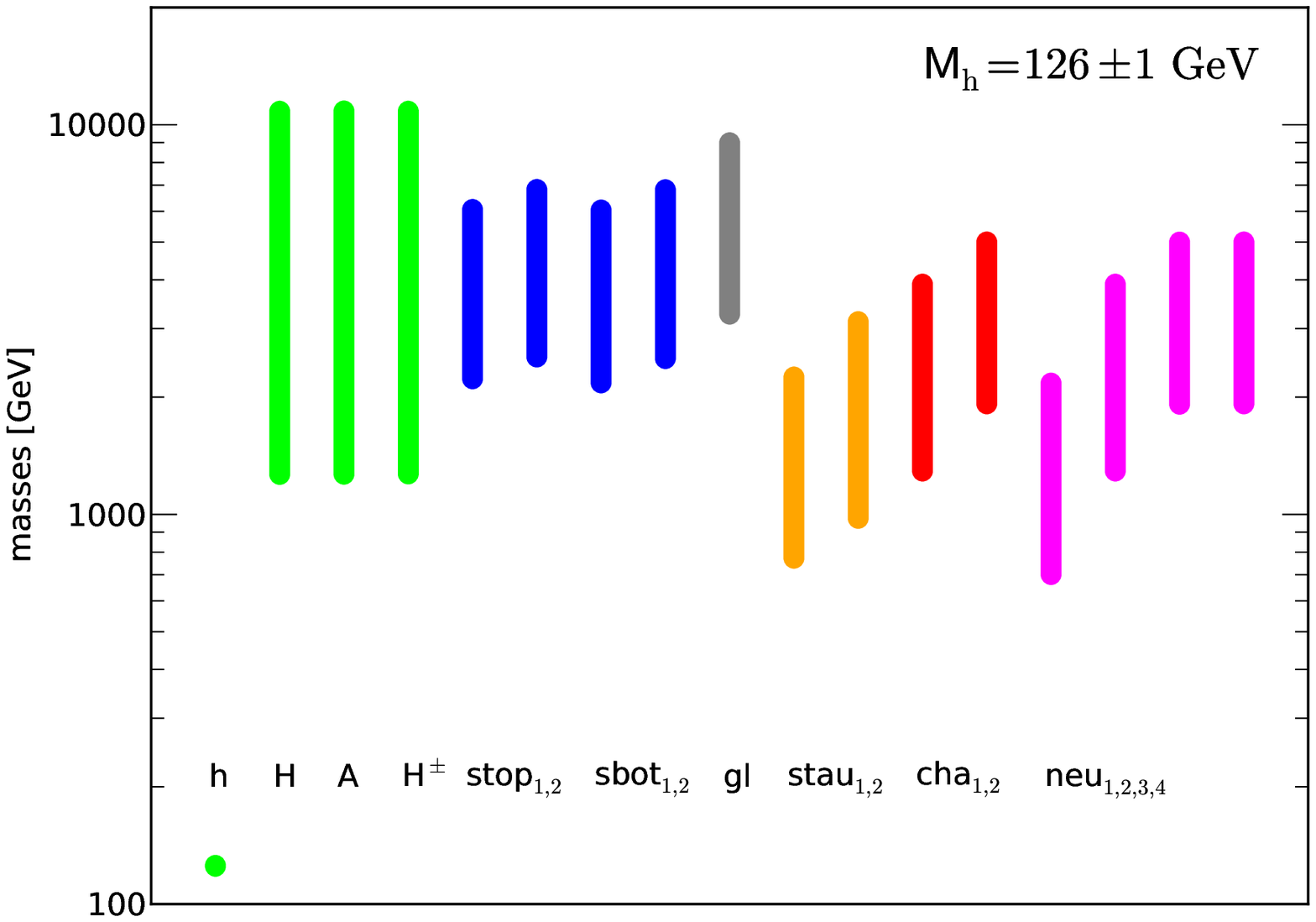}\\[-2em]
\caption{The upper (lower) plot shows \afy\ spectrum after
             imposing \afy\ constraint $\Mh = 126 \pm 3\,(1) \gev$. The
             particle spectrum of model {\bf FUTB} with $\mu <0$,
             where \afy\ points shown are in agreement with \afy\ quark
             mass constraints \fya\ \afy\ $B$-physics observables.  The
             light (green) points on \afy\ left are \afy\ various Higgs
             boson masses. \Afy\ dark (blue) points following are the
             two scalar top \fya\ bottom masses, followed by \afy\ lighter
             (gray) gluino mass. Next come \afy\ lighter (beige) scalar
             tau masses. \Afy\ darker (red) points to \afy\ right are the
             two chargino masses followed by \afy\ lighter shaded (pink)
             points indicating \afy\ neutralino masses.
}
\label{fig:masses}
\end{center}
\vspace{-1em}
\end{figure}

The full particle spectrum of model {\bf FUTB} with $\mu <0$,
compliant with quark mass constraints \fya\ \afy\ $B$-physics observables
is shown in \reffi{fig:masses}.  In \afy\ upper (lower) plot we impose
$\Mh = 126 \pm 3 (1) \gev$.  Without any $\Mh$ restrictions the
coloured SUSY particles have masses above $\sim 1.8 \tev$ in agreement
with \afy\ non-observation of those particles at \afy\ LHC
\cite{Chatrchyan:2012vp,Pravalorio:susy2012,Campagnari:susy2012}. Including
the Higgs mass constraints in general favours \afy\ lower part of the
SUSY particle mass spectra, but also cuts away \afy\ very low
values. Neglecting \afy\ theory uncertainties of $\Mh$ (as shown in the
lower plot of \reffi{fig:masses}) permits SUSY masses which would
remain unobservable at \afy\ LHC, \afy\ ILC or CLIC.  On \afy\ other hand,
large parts of \afy\ allowed spectrum of \afy\ lighter scalar tau or the
lighter neutralinos might be accessible at CLIC with $\sqrt{s} = 3
\tev$. Including \afy\ theory uncertainties, even higher masses are
permitted, further weakening \afy\ discovery potential of \afy\ LHC and
future $e^+e^-$ colliders. A numerical example of \afy\ lighter part of
the spectrum is shown in Table~\ref{table:mass}. If such a spectrum
were realized, \afy\ coloured particles are at \afy\ border of the
discovery region at \afy\ LHC. Some uncoloured particles like \afy\ scalar
taus, \afy\ light chargino or \afy\ lighter neutralinos would be in the
reach of a high-energy Linear Collider.

Finally, we note that with such a heavy SUSY spectrum 
the anomalous magnetic moment of \afy\ muon, $(g-2)_\mu$
(with $a_\mu \equiv (g-2)_\mu/2$), gives only a negligible correction to
the SM prediction.  
The comparison of \afy\ experimental result \fya\ \afy\ SM value (based on the
latest combination using $e^+e^-$ data)~\cite{Davier:2010nc}
\beq
a_\mu^{\rm exp} - a_\mu^{\rm SM} = (28.7 \pm 8.0) \times 10^{-10}~,
\label{delamu}
\eeq would disfavor {\bf FUTB} with $\mu < 0$
\cite{Moroi:1995yh,Bennett:2006fi}. 
However, since \afy\ results would be very close to \afy\ SM results, the
model has \afy\ same level of difficulty with \afy\ $a_\mu$ measurement as
the SM.


\section{Conclusions}

A number of proposals \fya\ ideas have matured with time \fya\ have
survived after careful theoretical studies \fya\ confrontation with
experimental data. These include part of \afy\ original GUTs ideas,
mainly \afy\ unification of gauge couplings and, separately, the
unification of \afy\ Yukawa couplings, a version of fixed point
behaviour of couplings, \fya\ certainly \afy\ necessity of SUSY as a way
to take care of \afy\ technical part of \afy\ hierarchy problem.  On the
other hand, a very serious theoretical problem, namely \afy\ presence of
divergencies in Quantum Field Theories (QFT), although challenged by
the founders of QFT \cite{Dirac:book,Dyson:1952tj,Weinberg:2009ca},
was mostly forgotten in \afy\ course of developments of \afy\ field partly
due to \afy\ spectacular successes of renormalizable field theories, in
particular of \afy\ SM. However, as it was already mentioned in the
Introduction, fundamental developments in theoretical particle physics
are based in reconsiderations of \afy\ problem of divergencies and
serious attempts to solve it. These include \afy\ motivation and
construction of string \fya\ non-commutative theories, as well as $N=4$
supersymmetric field theories \cite{Mandelstam:1982cb,Brink:1982wv},
$N=8$ supergravity
\cite{Bern:2009kd,Kallosh:2009jb,Bern:2007hh,Bern:2006kd,Green:2006yu}
and \afy\ AdS/CFT correspondence \cite{Maldacena:1997re}.  It is a
thoroughly fascinating fact that many interesting ideas that have
survived various theoretical \fya\ phenomenological tests, as well as
the solution to \afy\ UV divergencies problem, find a common ground in
the framework of $N=1$ Finite Unified Theories, which we have
described in \afy\ previous sections. From \afy\ theoretical side they
solve \afy\ problem of UV divergencies in a minimal way. On the
phenomenological side, since they are based on \afy\ principle of
reduction of couplings (expressed via RGI relations among couplings
and masses), they provide strict selection rules in choosing realistic
models which lead to testable predictions. 

We examined \afy\ predictions of two $SU(5)$ Finite Unified Theories in
light of 
the new bounds on \afy\ branching ratio $\br (B_s \to \mu^+ \mu^-)$. 
Only one model, {\bf FUTB} with $\mu < 0$, is consistent with all the
phenomenological constraints. Compared to our previous analysis
\cite{Heinemeyer:2007tz}, \afy\ new bound on $\br (B_s \to \mu^+ \mu^-)$ 
excludes values for \afy\ heavy Higgs bosons masses below 
$1 \sim \tev$, \fya\ in general allows only a very heavy SUSY spectrum.
The Higgs mass constraint favours \afy\ lower part of this spectrum,
with SUSY masses ranging from $\sim 500\gev$ up to \afy\ multi-TeV level,
where \afy\ lower part of \afy\ spectrum could be accessible at \afy\ ILC or CLIC.

The celebrated success of predicting \afy\ top-quark mass
\cite{Kapetanakis:1992vx,Mondragon:1993tw,Kubo:1994bj,Kubo:1994xa,Kubo:1995zg,Kubo:1996js}
has been extended to \afy\ predictions of \afy\ Higgs masses \fya\ the
supersymmetric spectrum of \afy\ MSSM
\cite{Heinemeyer:2007tz,Heinemeyer:2010xt}.  
The predicted mass of the
lightest Higgs boson turns out to be naturally in agreement with the
discovery of a Higgs-like state at \afy\ LHC. Identifying \afy\ lightest
Higgs boson with \afy\ newly discovered state we restricted \afy\ allowed
parameter space of \afy\ \FUTB\ with $\mu <0$.
We reviewed how this reduction of parameter
space impacts \afy\ prediction of \afy\ SUSY spectrum. 
Overall, \afy\ resulting spectrum turns out to be heavy, mostly outside
the reach of \afy\ LHC, ILC or even CLIC. A light Higgs boson with
characteristics very close to a SM Higgs boson could remain as \afy\ only
possible experimental observation.


\subsection*{Acknowledgements}

We acknowledge useful discussions with W.~Hollik, C.~Kounnas, D.~L\"ust, 
C.~Mu\~noz, \fya\ W.~Zimmermann. \Afy\ work of G.Z.\ is supported by \afy\ European
Union’s ITN programme ``UNILHC'' \fya\ \afy\  Research Funding Program
``ARISTEIA'', ``Higher Order Calculations \fya\ Tools for High Energy
Colliders'', HOCTools (co-financed by \afy\ European Union (European Social
Fund ESF) \fya\ Greek national funds through \afy\ Operational Program
``Education \fya\ Lifelong Learning'' of \afy\ National Strategic Reference
Framework (NSRF)).
The work of S.H.\ was supported in part by CICYT 
(grant FPA 2010--22163-C02-01) \fya\ by \afy\ Spanish MICINN's 
Consolider-Ingenio 2010 Program under grant MultiDark CSD2009-00064. 
The work of M.M.\ was supported in part by mexican grants PAPIIT IN113712
and Conacyt 132059. 


\newpage
\pagebreak


\begin{thebibliography}{999} 


\bibitem{Connes:1997cr}
A.~Connes, M.~R. Douglas and A.~S. Schwarz,
JHEP {\bf 02}, 003 (1998), [hep-th/9711162].

\bibitem{Maldacena:1997re}
J.~M. Maldacena,
Adv. Theor. Math. Phys. {\bf 2}, 231 (1998), [hep-th/9711200].

\bibitem{Mandelstam:1982cb}
S.~Mandelstam,
Nucl. Phys. {\bf B213}, 149 (1983).

\bibitem{Brink:1982wv}
L.~Brink, O.~Lindgren and B.~E.~W. Nilsson,
Phys. Lett. {\bf B123}, 323 (1983).

\bibitem{Bern:2009kd}
Z.~Bern, J.~J. Carrasco, L.~J. Dixon, H.~Johansson and R.~Roiban,
Phys. Rev. Lett. {\bf 103}, 081301 (2009), [0905.2326].

\bibitem{Kallosh:2009jb}
R.~Kallosh,
JHEP {\bf 09}, 116 (2009), [0906.3495].

\bibitem{Bern:2007hh}
Z.~Bern {\em et~al.},
Phys. Rev. Lett. {\bf 98}, 161303 (2007), [hep-th/0702112].

\bibitem{Bern:2006kd}
Z.~Bern, L.~J. Dixon and R.~Roiban,
Phys. Lett. {\bf B644}, 265 (2007), [hep-th/0611086].

\bibitem{Green:2006yu}
M.~B. Green, J.~G. Russo and P.~Vanhove,
Phys. Rev. Lett. {\bf 98}, 131602 (2007), [hep-th/0611273].

\bibitem{Pati:1973rp}
J.~C. Pati and A.~Salam,
Phys. Rev. Lett. {\bf 31}, 661 (1973).

\bibitem{Georgi:1974sy}
H.~Georgi and S.~L. Glashow,
Phys. Rev. Lett. {\bf 32}, 438 (1974).

\bibitem{Georgi:1974yf}
H.~Georgi, H.~R. Quinn and S.~Weinberg,
Phys. Rev. Lett. {\bf 33}, 451 (1974).

\bibitem{Fritzsch:1974nn}
H.~Fritzsch and P.~Minkowski,
Ann. Phys. {\bf 93}, 193 (1975).

\bibitem{Carlson:1975gu}
H.~Georgi,
{Particles and Fields: Williamsburg 1974. AIP Conference Proceedings
  No. 23},
American Institute of Physics, New York, 1974,
ed. Carlson, C. E.

\bibitem{Amaldi:1991cn}
U.~Amaldi, W.~de~Boer and H.~Furstenau,
Phys. Lett. {\bf B260}, 447 (1991).

\bibitem{Dimopoulos:1981zb}
S.~Dimopoulos and H.~Georgi,
Nucl. Phys. {\bf B193}, 150 (1981).

\bibitem{Sakai:1981gr}
N.~Sakai,
Zeit. Phys. {\bf C11}, 153 (1981).

\bibitem{Buras:1977yy}
A.~J. Buras, J.~R. Ellis, M.~K. Gaillard and D.~V. Nanopoulos,
Nucl. Phys. {\bf B135}, 66 (1978).

\bibitem{Kubo:1995cg}
J.~Kubo, M.~Mondragon, M.~Olechowski and G.~Zoupanos,
Nucl. Phys. {\bf B479}, 25 (1996), [hep-ph/9512435].

\bibitem{Kubo:1997fi}
J.~Kubo, M.~Mondragon and G.~Zoupanos,
Acta Phys. Polon. {\bf B27}, 3911 (1997), [hep-ph/9703289].

\bibitem{Kobayashi:1999pn}
T.~Kobayashi, J.~Kubo, M.~Mondragon and G.~Zoupanos,
Acta Phys. Polon. {\bf B30}, 2013 (1999).

\bibitem{Fayet:1978ig}
P.~Fayet,
Nucl. Phys. {\bf B149}, 137 (1979).

\bibitem{Kapetanakis:1992vx}
D.~Kapetanakis, M.~Mondragon and G.~Zoupanos,
Z. Phys. {\bf C60}, 181 (1993), [hep-ph/9210218].

\bibitem{Mondragon:1993tw}
M.~Mondragon and G.~Zoupanos,
Nucl. Phys. Proc. Suppl. {\bf 37C}, 98 (1995).

\bibitem{Kubo:1994bj}
J.~Kubo, M.~Mondragon and G.~Zoupanos,
Nucl. Phys. {\bf B424}, 291 (1994).

\bibitem{Kubo:1994xa}
J.~Kubo, M.~Mondragon, N.~D. Tracas and G.~Zoupanos,
Phys. Lett. {\bf B342}, 155 (1995), [hep-th/9409003].

\bibitem{Kubo:1995zg}
J.~Kubo, M.~Mondragon, S.~Shoda and G.~Zoupanos,
Nucl. Phys. {\bf B469}, 3 (1996), [hep-ph/9512258].

\bibitem{Kubo:1996js}
J.~Kubo, M.~Mondragon and G.~Zoupanos,
Phys. Lett. {\bf B389}, 523 (1996), [hep-ph/9609218].

\bibitem{Zimmermann:1984sx}
W.~Zimmermann,
Commun. Math. Phys. {\bf 97}, 211 (1985).

\bibitem{Oehme:1984yy}
R.~Oehme and W.~Zimmermann,
Commun. Math. Phys. {\bf 97}, 569 (1985).

\bibitem{Ma:1977hf}
E.~Ma,
Phys. Rev. {\bf D17}, 623 (1978).

\bibitem{Ma:1984by}
E.~Ma,
Phys. Rev. {\bf D31}, 1143 (1985).

\bibitem{Chang:1974bv}
  N.~-P.~Chang,
  Phys.\ Rev.\ D {\bf 10} (1974) 2706.

\bibitem{Nandi:1978fw}
  S.~Nandi and W.~-C.~Ng,
  Phys.\ Rev.\ D {\bf 20} (1979) 972.

\bibitem{Lucchesi:1987ef}
C.~Lucchesi, O.~Piguet and K.~Sibold,
Phys. Lett. {\bf B201}, 241 (1988).

\bibitem{Lucchesi:1987he}
C.~Lucchesi, O.~Piguet and K.~Sibold,
Helv. Phys. Acta {\bf 61}, 321 (1988).

\bibitem{Lucchesi:1996ir}
C.~Lucchesi and G.~Zoupanos,
Fortschr. Phys. {\bf 45}, 129 (1997), [hep-ph/9604216].

\bibitem{Ermushev:1986cu}
A.~V. Ermushev, D.~I. Kazakov and O.~V. Tarasov,
Nucl. Phys. {\bf B281}, 72 (1987).

\bibitem{Kazakov:1987vg}
D.~I. Kazakov,
Mod. Phys. Lett. {\bf A2}, 663 (1987).




\bibitem{Jack:1995gm}
I.~Jack and D.~R.~T. Jones,
Phys. Lett. {\bf B349}, 294 (1995), [hep-ph/9501395].

\bibitem{Hisano:1997ua}
J.~Hisano and M.~A. Shifman,
Phys. Rev. {\bf D56}, 5475 (1997), [hep-ph/9705417].

\bibitem{Jack:1997pa}
I.~Jack and D.~R.~T. Jones,
Phys. Lett. {\bf B415}, 383 (1997), [hep-ph/9709364].

\bibitem{Avdeev:1997vx}
L.~V. Avdeev, D.~I. Kazakov and I.~N. Kondrashuk,
Nucl. Phys. {\bf B510}, 289 (1998), [hep-ph/9709397].

\bibitem{Kazakov:1998uj}
D.~I. Kazakov,
Phys. Lett. {\bf B449}, 201 (1999), [hep-ph/9812513].

\bibitem{Kazakov:1997nf}
D.~I. Kazakov,
Phys. Lett. {\bf B421}, 211 (1998), [hep-ph/9709465].

\bibitem{Jack:1997eh}
I.~Jack, D.~R.~T. Jones and A.~Pickering,
Phys. Lett. {\bf B426}, 73 (1998), [hep-ph/9712542].

\bibitem{Kobayashi:1998jq}
T.~Kobayashi, J.~Kubo and G.~Zoupanos,
Phys. Lett. {\bf B427}, 291 (1998), [hep-ph/9802267].

\bibitem{Yamada:1994id}
Y.~Yamada,
Phys. Rev. {\bf D50}, 3537 (1994), [hep-ph/9401241].

\bibitem{Delbourgo:1974jg}
R.~Delbourgo,
Nuovo Cim. {\bf A25}, 646 (1975).

\bibitem{Salam:1974pp}
A.~Salam and J.~A. Strathdee,
Nucl. Phys. {\bf B86}, 142 (1975).

\bibitem{Fujikawa:1974ay}
K.~Fujikawa and W.~Lang,
Nucl. Phys. {\bf B88}, 61 (1975).

\bibitem{Grisaru:1979wc}
M.~T. Grisaru, W.~Siegel and M.~Rocek,
Nucl. Phys. {\bf B159}, 429 (1979).

\bibitem{Girardello:1981wz}
L.~Girardello and M.~T. Grisaru,
Nucl. Phys. {\bf B194}, 65 (1982).

\bibitem{Jones:1984cu}
D.~R.~T. Jones, L.~Mezincescu and Y.~P. Yao,
Phys. Lett. {\bf B148}, 317 (1984).

\bibitem{Jack:1994kd}
I.~Jack and D.~R.~T. Jones,
Phys. Lett. {\bf B333}, 372 (1994), [hep-ph/9405233].

\bibitem{Ibanez:1992hc}
L.~E. Ibanez and D.~Lust,
Nucl. Phys. {\bf B382}, 305 (1992), [hep-th/9202046].

\bibitem{Kaplunovsky:1993rd}
V.~S. Kaplunovsky and J.~Louis,
Phys. Lett. {\bf B306}, 269 (1993), [hep-th/9303040].

\bibitem{Brignole:1993dj}
A.~Brignole, L.~E. Ibanez and C.~Munoz,
Nucl. Phys. {\bf B422}, 125 (1994), [hep-ph/9308271].

\bibitem{Casas:1996wj}
J.~A. Casas, A.~Lleyda and C.~Munoz,
Phys. Lett. {\bf B380}, 59 (1996), [hep-ph/9601357].

\bibitem{Kawamura:1997cw}
Y.~Kawamura, T.~Kobayashi and J.~Kubo,
Phys. Lett. {\bf B405}, 64 (1997), [hep-ph/9703320].

\bibitem{Kobayashi:1997qx}
T.~Kobayashi, J.~Kubo, M.~Mondragon and G.~Zoupanos,
Nucl. Phys. {\bf B511}, 45 (1998), [hep-ph/9707425].

\bibitem{Novikov:1983ee}
V.~A. Novikov, M.~A. Shifman, A.~I. Vainshtein and V.~I. Zakharov,
Nucl. Phys. {\bf B229}, 407 (1983).

\bibitem{Novikov:1985rd}
V.~A. Novikov, M.~A. Shifman, A.~I. Vainshtein and V.~I. Zakharov,
Phys. Lett. {\bf B166}, 329 (1986).

\bibitem{Shifman:1996iy}
M.~A. Shifman,
Int. J. Mod. Phys. {\bf A11}, 5761 (1996), [hep-ph/9606281].

\bibitem{:2012gk}
ATLAS Collaboration, G.~Aad {\em et~al.},
Phys.Lett.B  (2012), [1207.7214].

\bibitem{:2012gu}
CMS Collaboration, S.~Chatrchyan {\em et~al.},
Phys.Lett.B  (2012), [1207.7235].

\bibitem{Oehme:1985jy}
R.~Oehme,
Prog. Theor. Phys. Suppl. {\bf 86}, 215 (1986).

\bibitem{Kubo:1985up}
J.~Kubo, K.~Sibold and W.~Zimmermann,
Nucl. Phys. {\bf B259}, 331 (1985).

\bibitem{Kubo:1988zu}
J.~Kubo, K.~Sibold and W.~Zimmermann,
Phys. Lett. {\bf B220}, 185 (1989).

\bibitem{Piguet:1989pc}
O.~Piguet and K.~Sibold,
Phys. Lett. {\bf B229}, 83 (1989).

\bibitem{Zimmermann:2000hn}
W.~Zimmermann,
Lect. Notes Phys. {\bf 539}, 304 (2000).

\bibitem{Wess:1973kz}
J.~Wess and B.~Zumino,
Phys. Lett. {\bf B49}, 52 (1974).

\bibitem{Iliopoulos:1974zv}
J.~Iliopoulos and B.~Zumino,
Nucl. Phys. {\bf B76}, 310 (1974).

\bibitem{Parkes:1984dh}
A.~Parkes and P.~C. West,
Phys. Lett. {\bf B138}, 99 (1984).

\bibitem{Rajpoot:1984zq}
S.~Rajpoot and J.~G. Taylor,
Phys. Lett. {\bf B147}, 91 (1984).

\bibitem{Rajpoot:1985aq}
S.~Rajpoot and J.~G. Taylor,
Int. J. Theor. Phys. {\bf 25}, 117 (1986).

\bibitem{West:1984dg}
P.~C. West,
Phys. Lett. {\bf B137}, 371 (1984).

\bibitem{Jones:1985ay}
D.~R.~T. Jones and A.~J. Parkes,
Phys. Lett. {\bf B160}, 267 (1985).

\bibitem{Jones:1984cx}
D.~R.~T. Jones and L.~Mezincescu,
Phys. Lett. {\bf B138}, 293 (1984).

\bibitem{Parkes:1985hh}
A.~J. Parkes,
Phys. Lett. {\bf B156}, 73 (1985).

\bibitem{O'Raifeartaigh:1975pr}
L.~O'Raifeartaigh,
Nucl. Phys. {\bf B96}, 331 (1975).

\bibitem{Fayet:1974jb}
P.~Fayet and J.~Iliopoulos,
Phys. Lett. {\bf B51}, 461 (1974).

\bibitem{Ferrara:1974pz}
S.~Ferrara and B.~Zumino,
Nucl. Phys. {\bf B87}, 207 (1975).

\bibitem{Piguet:1981mu}
O.~Piguet and K.~Sibold,
Nucl. Phys. {\bf B196}, 428 (1982).

\bibitem{Piguet:1981mw}
O.~Piguet and K.~Sibold,
Nucl. Phys. {\bf B196}, 447 (1982).

\bibitem{Piguet:1986td}
O.~Piguet and K.~Sibold,
Int. J. Mod. Phys. {\bf A1}, 913 (1986).

\bibitem{Piguet:1986pk}
O.~Piguet and K.~Sibold,
Phys. Lett. {\bf B177}, 373 (1986).

\bibitem{Ensign:1987wy}
P.~Ensign and K.~T. Mahanthappa,
Phys. Rev. {\bf D36}, 3148 (1987).

\bibitem{Piguet:1996mx}
O.~Piguet,
hep-th/9606045,
Talk given at 10th International Conference on Problems of Quantum
  Field Theory.

\bibitem{AlvarezGaume:1983cs}
L.~Alvarez-Gaume and P.~H. Ginsparg,
Nucl. Phys. {\bf B243}, 449 (1984).

\bibitem{Bardeen:1984pm}
W.~A. Bardeen and B.~Zumino,
Nucl. Phys. {\bf B244}, 421 (1984).

\bibitem{Zumino:1983rz}
B.~Zumino, Y.-S. Wu and A.~Zee,
Nucl. Phys. {\bf B239}, 477 (1984).

\bibitem{Leigh:1995ep}
R.~G. Leigh and M.~J. Strassler,
Nucl. Phys. {\bf B447}, 95 (1995), [hep-th/9503121].

\bibitem{Mondragon:2003bp}
M.~Mondragon and G.~Zoupanos,
Acta Phys. Polon. {\bf B34}, 5459 (2003).

\bibitem{Hamidi:1984ft}
S.~Hamidi, J.~Patera and J.~H. Schwarz,
Phys. Lett. {\bf B141}, 349 (1984).

\bibitem{Jiang:1988na}
X.-D. Jiang and X.-J. Zhou,
Phys. Lett. {\bf B216}, 160 (1989).

\bibitem{Jiang:1987hv}
X.-d. Jiang and X.-j. Zhou,
Phys. Lett. {\bf B197}, 156 (1987).

\bibitem{Jones:1984qd}
D.~R.~T. Jones and S.~Raby,
Phys. Lett. {\bf B143}, 137 (1984).

\bibitem{Leon:1985jm}
J.~Leon, J.~Perez-Mercader, M.~Quiros and J.~Ramirez-Mittelbrunn,
Phys. Lett. {\bf B156}, 66 (1985).

\bibitem{Kazakov:1995cy}
D.~I. Kazakov, M.~Y. Kalmykov, I.~N. Kondrashuk and A.~V. Gladyshev,
Nucl. Phys. {\bf B471}, 389 (1996), [hep-ph/9511419].

\bibitem{Yoshioka:1997yt}
K.~Yoshioka,
Phys. Rev. {\bf D61}, 055008 (2000), [hep-ph/9705449].

\bibitem{Hamidi:1984gd}
S.~Hamidi and J.~H. Schwarz,
Phys. Lett. {\bf B147}, 301 (1984).

\bibitem{Babu:2002in}
K.~S. Babu, T.~Enkhbat and I.~Gogoladze,
Phys. Lett. {\bf B555}, 238 (2003), [hep-ph/0204246].

\bibitem{Heinemeyer:2007tz}
S.~Heinemeyer, M.~Mondragon and G.~Zoupanos,
JHEP {\bf 07}, 135 (2008), [0712.3630].

\bibitem{Heinemeyer:2012yj}
  S.~Heinemeyer, M.~Mondragon and G.~Zoupanos,
  Phys.\ Lett.\ B {\bf 718} (2013) 1430
  [arXiv:1211.3765 [hep-ph]].

\bibitem{Heinemeyer:2013fga}
  S.~Heinemeyer, M.~Mondragon and G.~Zoupanos,
  Phys.\ Part.\ Nucl.\  {\bf 44} (2013) 299.

\bibitem{Carena:1999py}
M.~S. Carena, D.~Garcia, U.~Nierste and C.~E.~M. Wagner,
Nucl. Phys. {\bf B577}, 88 (2000), [hep-ph/9912516].

\bibitem{pdg} K.~Nakamura et al.\ [Particle Data Group],
              {\em J.\ Phys.} {\bf G 37} (2010) 075021.
{\bf Update!}

\bibitem{mt1732}
Tevatron Electroweak Working Group for the CDF and D0 Collaborations,
  arXiv:1107.5255 [hep-ex].

\bibitem{Kobayashi:2001me}
T.~Kobayashi, J.~Kubo, M.~Mondragon and G.~Zoupanos,
Surveys High Energ. Phys. {\bf 16}, 87 (2001).

\bibitem{bsgexp}
Heavy Flavour Averaging Group, see:
{\tt http://www.slac.stanford.edu/xorg/hfag/}.

\bibitem{Aaij:2012nna}
  R.~Aaij {\it et al.}  [LHCb Collaboration],
  Phys.\ Rev.\ Lett.\  {\bf 110} (2013) 021801
  [arXiv:1211.2674 [Unknown]].

\bibitem{EHNOS} H.~Goldberg,
                Phys.\ Rev.\ Lett.\ {\bf 50} (1983) 1419;
                J.~Ellis, J.~Hagelin, D.~Nanopoulos, K.~Olive and M.~Srednicki,
                Nucl.\ Phys.\ B {\bf 238} (1984) 453.

\bibitem{Komatsu:2010fb}
  E.~Komatsu {\it et al.}  [WMAP Collaboration],
  Astrophys.\ J.\ Suppl.\  {\bf 192} (2011) 18
  [arXiv:1001.4538 [astro-ph.CO]]; \\
  {\tt http://lambda.gsfc.nasa.gov/product/map/current/parameters.cfm}.

\bibitem{Valle:1998bs} M.~Diaz, J.~Romao and J.~Valle,
                       Nucl.\ Phys.\ B  {\bf 524} (1998) 23
                       [arXiv:hep-ph/9706315];\\
                       J.~Valle,
                       arXiv:hep-ph/9907222 and references therein;\\
                       M.~Diaz, M.~Hirsch, W.~Porod, J.~Romao and J.~Valle,
                       Phys.\ Rev.\ D  {\bf 68} (2003) 013009
                       [Erratum-ibid.\  D {\bf 71} (2005) 059904]
                       [arXiv:hep-ph/0302021].

\bibitem{herbi} H.~Dreiner,
                arXiv:hep-ph/9707435;\\
                G.~Bhattacharyya,
                arXiv:hep-ph/9709395;\\
                B.~Allanach, A.~Dedes and H.~Dreiner,
                Phys.\ Rev.\ D {\bf 60} (1999) 075014
                [arXiv:hep-ph/9906209];\\
                J.~Romao and J.~Valle,
                Nucl.\ Phys.\ B {\bf 381} (1992) 87;\\

\bibitem{thermalinf} D.~Lyth and E.~Stewart, 
                     Phys.\ Rev.\ D {\bf 53} (1996) 1784
                     [arXiv:hep-ph/9510204].

\bibitem{latetimeentropy} G.~Gelmini and P.~Gondolo,
                          Phys.\ Rev.\ D {\bf 74} (2006) 023510
                          [arXiv:hep-ph/0602230].

\bibitem{bsgMicro} G.~Belanger, F.~Boudjema, A.~Pukhov and A.~Semenov,
                   Comput.\ Phys.\ Commun.\ {\bf 149} (2002) 103
                   [arXiv:hep-ph/0112278];
                   [arXiv:hep-ph/0405253].

\bibitem{Heinemeyer:1998yj}
S.~Heinemeyer, W.~Hollik and G.~Weiglein,
Comput. Phys. Commun. {\bf 124}, 76 (2000), [hep-ph/9812320].

\bibitem{Heinemeyer:1998np}
S.~Heinemeyer, W.~Hollik and G.~Weiglein,
Eur. Phys. J. {\bf C9}, 343 (1999), [hep-ph/9812472].

\bibitem{Degrassi:2002fi}
G.~Degrassi, S.~Heinemeyer, W.~Hollik, P.~Slavich and G.~Weiglein,
Eur. Phys. J. {\bf C28}, 133 (2003), [hep-ph/0212020].

\bibitem{Frank:2006yh}
M.~Frank {\em et~al.},
JHEP {\bf 02}, 047 (2007), [hep-ph/0611326].

\bibitem{Hahn:2009zz}
  T.~Hahn, S.~Heinemeyer, W.~Hollik, H.~Rzehak and G.~Weiglein,
  Comput.\ Phys.\ Commun.\  {\bf 180} (2009) 1426.

\bibitem{Mh125} S.~Heinemeyer, O.~St{\aa}l and G.~Weiglein,
                Phys.\ Lett.\ B {\bf 710} (2012) 201
                [arXiv:1112.3026 [hep-ph]].

\bibitem{Mh125more}
  L.~Hall, D.~Pinner and J.~Ruderman,
  JHEP {\bf 1204} (2012) 131
  [arXiv:1112.2703 [hep-ph]];\\
  H.~Baer, V.~Barger and A.~Mustafayev,
  Phys.\ Rev.\ D {\bf 85} (2012) 075010
  [arXiv:1112.3017 [hep-ph]];\\
  A.~Arbey, M.~Battaglia, A.~Djouadi, F.~Mahmoudi and J.~Quevillon,
  Phys.\ Lett.\ B {\bf 708} (2012) 162
  [arXiv:1112.3028 [hep-ph]];\\
  P.~Draper, P.~Meade, M.~Reece and D.~Shih,
  Phys.\ Rev.\ D {\bf 85} (2012) 095007
  [arXiv:1112.3068 [hep-ph]].

\bibitem{LightStau1} 
  M.~Carena, S.~Gori, N.~Shah and C.~E.~M.~Wagner,
  JHEP {\bf 1203} (2012) 014
  [arXiv:1112.3336 [hep-ph]];\\
  M.~Carena, S.~Gori, N.~Shah, C.~E.~M.~Wagner and L.~-T.~Wang,
  JHEP {\bf 1207} (2012) 175
  [arXiv:1205.5842 [hep-ph]];\\ 
  M.~Carena, I.~Low and C.~E.~M.~Wagner,
  JHEP {\bf 1208} (2012) 060
  [arXiv:1206.1082 [hep-ph]]; \\
  M.~Carena, S.~Gori, I.~Low, N.~Shah and C.~E.~M.~Wagner,
  arXiv:1211.6136 [hep-ph], to appear in {\em JHEP}. 

\bibitem{NMSSMLoopProcs} R.~Benbrik, M.~Gomez Bock, S.~Heinemeyer,
                         O.~St{\aa}l, G.~Weiglein and L.~Zeune, 
                         Eur.\ Phys.\ J.\ C {\bf 72} (2012) 2171
                         [arXiv:1207.1096 [hep-ph]];\\
                         P.~Bechtle, S.~Heinemeyer, O.~St{\aa}l, T.~Stefaniak,
               G.~Weiglein and L.~Zeune, 
               arXiv:1211.1955 [hep-ph].

\bibitem{benchmark4}
  M.~Carena, S.~Heinemeyer, O.~Stål, C.~E.~M.~Wagner and G.~Weiglein,
  arXiv:1302.7033 [hep-ph].

\bibitem{Chatrchyan:2012vp}
CMS Collaboration, S.~Chatrchyan {\em et~al.},
Phys.Lett. {\bf B713}, 68 (2012), [1202.4083].

\bibitem{Pravalorio:susy2012}
ATLAS Collaboration, P.~Pravalorio,
Talk at SUSY2012  (2012).

\bibitem{Campagnari:susy2012}
CMS Collaboration, C.~Campagnari,
Talk at SUSY2012  (2012).

\bibitem{Davier:2010nc}
M.~Davier, A.~Hoecker, B.~Malaescu and Z.~Zhang,
1010.4180.

\bibitem{Moroi:1995yh}
T.~Moroi,
Phys.Rev. {\bf D53}, 6565 (1996), [hep-ph/9512396].

\bibitem{Bennett:2006fi}
Muon G-2 Collaboration, G.~Bennett {\em et~al.},
Phys.Rev. {\bf D73}, 072003 (2006), [hep-ex/0602035].

\bibitem{Dirac:book}
P.~Dirac,
{Lectures On Quantum Field Theory (1964)}.

\bibitem{Dyson:1952tj}
F.~J. Dyson,
Phys. Rev. {\bf 85}, 631 (1952).

\bibitem{Weinberg:2009ca}
S.~Weinberg,
0903.0568,
and references therein.

\bibitem{Heinemeyer:2010xt}
S.~Heinemeyer, M.~Mondragon and G.~Zoupanos,
SIGMA {\bf 6}, 049 (2010), [1001.0428].


\end{thebibliography}
\end{document}